\documentclass[12pt,onecolumn,draftclsnofoot]{IEEEtran}
\IEEEoverridecommandlockouts
\usepackage[left=1in,top=1.2in,right=1in,bottom=1.2in]{geometry}
\usepackage{mathtools,xparse}
\usepackage{amsmath,amssymb,amsfonts}
\usepackage{geometry}

\usepackage{amsthm}
\usepackage{float}
\usepackage[caption=false]{subfig}

\usepackage{cite}
\usepackage{relsize}

\usepackage{float}
\usepackage{slashbox}
\usepackage{mathtools}
\usepackage{graphics}
\usepackage{bm}
\usepackage{amsmath}
\usepackage{algorithmic}
\usepackage{graphicx}
\usepackage{textcomp}
\usepackage{xcolor}
\usepackage{etoolbox}
\usepackage{lipsum}

\let\mybibitem\bibitem
\renewcommand{\bibitem}[1]{%
\ifstrequal{#1}{joung2010multiuser}{\color{black}\mybibitem{#1}}
{\ifstrequal{#1}{marzetta2016fundamentals}{\color{black}\mybibitem{#1}}
{\ifstrequal{#1}{zhang2018spectral}{\color{black}\mybibitem{#1}}
{\ifstrequal{#1}{ong2011capacity}{\color{black}\mybibitem{#1}}
{\ifstrequal{#1}{bjornson2017massive}{\color{black}\mybibitem{#1}}
{\ifstrequal{#1}{zhang2018low}{\color{black}\mybibitem{#1}}	
{\ifstrequal{#1}{lee2008analog}{\color{black}\mybibitem{#1}}	
{\ifstrequal{#1}{zhang2016spectral2}{\color{black}\mybibitem{#1}}	
{\ifstrequal{#1}{li2017channel}{\color{black}\mybibitem{#1}}		
{\ifstrequal{#1}{liang2016mixed}{\color{black}\mybibitem{#1}}	
{\ifstrequal{#1}{li2016channel}{\color{black}\mybibitem{#1}}	
{\ifstrequal{#1}{mezghani2012capacity}{\color{black}\mybibitem{#1}}		
{\ifstrequal{#1}{dong2017efficient}{\color{black}\mybibitem{#1}}		
{\ifstrequal{#1}{amah2010beamforming}{\color{black}\mybibitem{#1}}
{\ifstrequal{#1}{studer2016quantized}{\color{black}\mybibitem{#1}}
{\ifstrequal{#1}{jacobsson2015one}{\color{black}\mybibitem{#1}}	
{\ifstrequal{#1}{choi2016near}{\color{black}\mybibitem{#1}}	
{\ifstrequal{#1}{mo2018channel}{\color{black}\mybibitem{#1}}	
{\color{black}\mybibitem{#1}}}}}}}}}}}}}}}}}}}
}%

\DeclarePairedDelimiter{\norm}{\lVert}{\rVert}
\DeclarePairedDelimiter{\abs}{\lvert}{\rvert}
\newtheorem{Theo.}{Theorem}
\newtheorem{Lemma}{Lemma}

\def\bD{{\bf D}}

\begin{document}

\makeatletter
\def\blfootnote{\xdef\@thefnmark{}\@footnotetext}
\makeatother
        %
        \title{Performance Analysis of Massive MIMO Multi-Way Relay Networks with Low-Resolution ADCs}

        \author{\IEEEauthorblockN{Samira Rahimian, Yindi Jing, Masoud Ardakani\\ }
                \IEEEauthorblockA{ University of Alberta, Canada}}
        \maketitle
 \vspace{-2.5cm}
        \textcolor{black} {\begin{abstract}
       	\blfootnote{\color{black} Part of this work on the performance analysis of mMIMO multi-way relay networks with low-resolution uniform-ADC structures has been presented at the IEEE International Conference on Communications (ICC) 2019 \cite{Rahi1905:Performance}.}High power consumption and hardware cost are two barriers for practical massive multiple-input multiple-output
       	(mMIMO) systems. A promising solution is to employ
       	low-resolution analog-to-digital converters (ADCs). In this paper, we consider a general mMIMO multi-way relaying system with a multi-level mixed-ADC architecture, in which each antenna is connected to an ADC pair of an arbitrary resolution. By leveraging on Bussgang's decomposition theorem and Lloyd-Max algorithm for quantization, tight closed-form approximations are derived for the average achievable rates of zero-forcing (ZF) relaying considering both perfect and imperfect channel state information (CSI). To conquer the challenges caused by multi-way relaying, the complicated ZF beam-forming matrix, and the general mixed-ADC structure, we develop a novel method for the achievable rate analysis using the singular-value decomposition (SVD) for Gaussian matrices, distributions of the singular values of Gaussian matrices, and properties of Haar matrices. The results explicitly show the achievable rate behavior in terms of the user and relay transmit powers and the numbers of relay antennas and users. Most importantly, it quantifies the performance degradation caused by low-resolution ADCs and channel estimation error.  We demonstrate that the average achievable rate has an almost linear relation with the square of the average of quantization coefficients pertaining to the ADC resolution profile. In addition, in the medium to high SNR region, the ADC resolutions have more significant effect on the rate compared to the number of antennas. Our work reveals that the performance gap between the perfect and imperfect CSI cases increases as the average ADC resolution increases. For more insights, simplified achievable rate expressions for asymptotic cases and the uniform-ADC case are obtained. Numerical results verify that the theoretical results can accurately predict the performance of the considered system.  
        	\end{abstract}}
            
      \vspace{-0.4cm}
        \begin{IEEEkeywords}
                Massive MIMO, multi-way communications,  low-resolution ADC, mixed-ADC, uniform-ADC, achievable rate, imperfect channel state information, asymptotic analysis
        \end{IEEEkeywords}

        %
        \IEEEpeerreviewmaketitle

        \section{Introduction}
\textcolor{black}{As one of the key technologies for the fifth generation (5G) of wireless communications, massive multi-input multi-output (mMIMO) has attracted extensive research interests in recent years \cite{andrews2014will,marzetta2016fundamentals}. By exploiting
	quasi-orthogonal random channel vectors between different
	users, mMIMO can mitigate the inter-user interference to
	provide high spectral and energy efficiency via simple linear signal processing, e.g., maximum-ratio combining (MRC) and zero-forcing (ZF). On the other hand, relaying is an important way of extending coverage and improving service. In multi-way relay networks (MWRNs) multiple interfering users communicate simultaneously to exchange messages such that each user multi-casts its message to all other users.  Compared to one-way and two-way relaying, multi-way relaying \cite{zhang2018spectral,gunduz2013multiway,ong2011capacity,amah2009non} can significantly reduce the number of time slots for full mutual communications among users which consequently improves the spectral and energy efficiencies. Hence, mMIMO MWRNs, where the relays are equipped with large-scale antenna arrays, benefit from the advantages of both multi-way relaying and mMIMO structures. For mMIMO MWRNs with ZF processing, \cite{amarasuriya2014multi,ho2017performance} have obtained closed-form approximations for the spectral and energy efficiencies. It is concluded that the transmit power of each user and the relay can be made inversely proportional to the number of relay antennas while maintaining required quality of service.}

\textcolor{black}{The practical implementation of mMIMO systems with large-scale antenna arrays is challenged by the high hardware cost and energy consumption \cite{bjornson2017massive,zhang2018low}. Typically, each receive and transmit antenna is connected to an analog-to-digital converter (ADC) and a digital-to-analog converter (DAC) in the radio frequency (RF) chain, respectively. Compared to mMIMO systems with all high-resolution ADCs and DACs (e.g., 8-12 bits), it is less costly and more energy efficient to employ low-cost, low-power, low-resolution ADCs and DACs (e.g., 1-4 bits) \cite{fan2015uplink,zhang2016spectral2}. Especially, the hardware cost and power consumption of ADCs grow exponentially with the number of quantization bits \cite{lee2008analog}. Naturally, signal processing challenges and complex front-end designs occur due to the nonlinear characteristic of coarse quantization \cite{li2017channel,gokceoglu2017spatio}.}
	
	 {\color{black}As this work considers the ADCs only, in what follows, we review the literature on mMIMO with low-resolution ADCs.
	 	 The primary works on this topic have considered that all ADCs have the same resolution, also referred to as uniform-ADC \cite{fan2015uplink,zhang2016spectral2,li2017channel,gokceoglu2017spatio,dong2017efficient,studer2016quantized}. For instance, considering frequency-selective channels, uplink performance of an mMIMO system with uniform-ADC that deploys orthogonal frequency-division multiplexing (OFDM) is investigated in \cite{studer2016quantized}, where new algorithms for quantized maximum a-posteriori channel estimation and data detection are proposed. It is shown that coarse quantization (e.g., 4-6 bits) in mMIMO-OFDM systems entails no performance loss compared with the full-resolution case.
Later, a two-level mixed-ADC architecture is proposed, in which part of the antennas are connected to low-resolution
	ADCs with the same resolution (usually 1-bit), while the remaining are connected to high-resolution (usually the ideal infinite-resolution) ADCs \cite{zhang2018spectral2,liang2016mixed}. In \cite{zhang2018spectral2}, the achievable uplink spectral efficiency of a mMIMO system with two-level mixed-ADC receiver assuming perfect CSI is investigated for MRC detector in the multi-cell scenario and ZF detector in the single-cell system. Further, for the two-level ADC structure the channel state information (CSI) obtainment schemes are proposed in \cite{liang2016mixed,jacobsson2015one,choi2016near,mo2018channel,pirzadeh2018spectral}, for example, by using the high-resolution ADCs in a round-robin manner \cite{liang2016mixed}. Recently, a general multi-level mixed-ADC structure is proposed in \cite{ding2018receiver} that allows multiple ADC levels and arbitrary ADC resolution profile for the large-scale antenna array. It provides more degrees-of-freedom compared to the two-level mixed-ADC and uniform-ADC architectures in achieving the desirable balance between performance and (hardware and energy) cost.} 
\vspace{-0.5cm}
{\color{black}\subsection{Relevant Prior Work}}
 \textcolor{black}{There have been many recent works on single-hop mMIMO systems with a mixed-ADC architecture. Among them, the mutual information and the spectral efficiency for the uplink of two-level mixed-ADC systems are investigated in \cite{liang2016mixed} and \cite{tan2016spectral}, respectively. For the uplink of mMIMO systems with multi-level mixed-ADC architecture and MRC processing, closed-form approximations for the spectral efficiency, receive energy efficiency, and outage probability are derived in \cite{ding2018receiver} and \cite{ding2018outage}. Also, these works study the optimization of ADC resolutions with certain goals on the achievable sum-rate, outage probability, and receive energy efficiency. These contributions have shown that the power consumption and hardware cost of the single-hop mMIMO system with a mixed-ADC architecture can be considerably
 	reduced while maintaining most of the gains in the achievable rate.} 
     
\textcolor{black}{There are a few research results on the two-hop mMIMO relaying system with low-resolution ADCs. Among them, \cite{kong2017full,liu2017multiuser,kong2018multipair,zhang2019mixed} have investigated the performance of multi-pair mMIMO one-way relaying systems.
In \cite{kong2017full}, the relay and all the users are assumed to have uniform-ADC where closed-form expressions for the achievable sum-rate are derived considering imperfect CSI and MRC/maximum-ratio transmission (MRT) processing at the relay. It is shown that with only low-resolution ADCs at the relay, increasing the number of relay antennas is effective to compensate for the rate loss caused by coarse quantization. However, it becomes ineffective to handle the detrimental effect of low-resolution ADCs at the users. Further, for mMIMO one-way relay systems with two-level mixed-ADC and MRC detection, the achievable rate is investigated in \cite{liu2017multiuser}, where it is shown that the performance loss due to the low-resolution ADCs can be compensated by increasing the number of relay antennas. The work in \cite{kong2018multipair} and \cite{zhang2019mixed} are on mMIMO one-way relay systems with both low-resolution ADCs and low-resolution DACs under CSI error and maximum ratio (MR) processing. In \cite{kong2018multipair}, for the case of uniform 1-bit ADCs and DACs, a closed-form asymptotic approximation for the achievable rate is derived. For the two-level mixed-ADCs and mixed-DACs, the work in \cite{zhang2019mixed} has derived exact and approximate closed-form expressions for the achievable rate.} {\color{black} The trade-off between the achievable rate and power consumption for different numbers of low-resolution ADCs/DACs is also investigated.}

 {\color{black}\subsection{Contributions}}
\textcolor{black}{To the best of our knowledge, performance analysis of mMIMO relaying systems with low-resolution ADCs has mainly focused on one-way relaying with uniform-ADC and two-level mixed-ADC, and there has been no result on the general multi-level mixed-ADC structure. {\color{black} Compared to uniform-ADC and two-level mixed-ADC profiles, the multi-level mixed-ADC structure is more general and provides the system designers with extra degrees-of-freedom for the design and optimization of  the system. For example, it enables the achievement of many more optimal points on the trade-off between the achievable rate and energy consumption, as shown in Figure 2 in \cite{ding2018receiver} and Figure 5 in \cite{ding2018outage}.} {\color{black} On the other hand, this general assumption imposes extra complication in performance analysis where existing methods cannot be applied directly. Further, there has been no work on the performance analysis of multi-level mixed-ADC structure with ZF beam-forming even for single-hop (uplink) scenarios, as previous studies are limited to MR processing \cite{ding2018receiver,ding2018outage}.}
		\textcolor{black}{The complicated ZF beam-forming brings further challenges in the analysis.} {\color{black}
		Moreover, compared to uplink communications and one-way relaying systems that were studied before, the multi-way relaying further complicates the performance analysis through the following aspects. 1) It has multiple broadcast time slots, each having a distinct ZF beam-forming matrix; and 2) the channel of the multiple access phase is the transpose of the channel of the broadcast phase, causing more contamination among different time slots.}}
	 	 
{\color{black}In this paper, for the first time, we derive the average achievable rate for mMIMO multi-way relay systems with a general multi-level mixed-ADC receiver. Both perfect and imperfect CSI cases are investigated. Further, ZF beam-forming is assumed at the relay as one of the most popular beam-forming designs that brings high performance in MWRNs, especially for the high SNR range \cite{ho2017performance}. The main contributions of this work are summarized as follows:}
  {\color{black}\begin{itemize}
 	\item To derive the ADC quantization model, Bussgang's decomposition theorem [32] is adopted to find the uncorrelated quantization noise to the quantization input. This is different from the additive quantization noise model (AQNM) that roughly models the quantization noise as an independent signal to the quantization input with Gaussian distribution. Furthermore, our work uses the mean squared error (MSE)-optimal sets of quantization labels and thresholds obtained from Lloyd-Max algorithm for the quantization.    
 	\item While existing derivation methods for mMIMO systems cannot be applied directly for the multi-way network under ZF, in this work, we develop a new method by firstly using singular-value decomposition (SVD) for Gaussian matrices to simplify the expressions, then applying properties of Wishart distributed matrices, Haar (isotropically distributed) matrices, and the distribution of singular values for the Gaussian matrices with i.i.d. entries. This novel method enables us to find a tight expression for the average achievable rates and can be applied to other similar scenarios  under ZF beam-forming.  The proposed method is fundamentally different from the truncation-based approximation in [19] and truncation error is avoided. Further, it can be applied in other systems with ZF beam-forming or other beam-forming schemes with a similar structure.
 	
 	\item The obtained results provide insights into the effects of user transmit power, relay transmit power, the number of relay antennas, the number of users, channel estimation error and most importantly the ADC resolution profile  on the achievable rate. It is shown that in the medium to high SNR region, the ADC resolutions have more significant effect on the rate compared to the number of antennas. 

 	\item For two asymptotic cases, simplified expressions are derived for the average achievable rates. One case is when the number of relay antennas approaches infinity while the number of users is fixed. The result in this case reveals a linear relationship between the average achievable rate and the number of antennas at the relay. The other case is when the numbers of users and relay antennas increase toward infinity with a fixed ratio, referred to as the loading factor. The result proves an inverse linear relationship between the achievable rate and the loading factor. In addition, we have provided results for the special case of uniform-ADC which shows that the average achievable rate has a linear relation with the square of the average of the quantization coefficients pertaining to the ADC resolution. Also, the square of the average of the quantization coefficients and the user power always appear together and can compensate for each other.
 	\item The achievable rate analysis is extended to the imperfect CSI case where a closed-form approximation is derived. It is shown that the gap between the achievable rates for perfect and imperfect CSI cases gets larger as the average of the ADC resolutions increases. This inspires that for practical systems with limited CSI quality, using lower resolution ADCs can gain significantly better energy efficiency and hardware cost while maintaining most of the rate performance.
 \end{itemize}}
\vspace{-.5cm}
{ \color{black} \subsection{Paper Outline and Notations}}
 The rest of this paper is organized as follows. Our system model is presented in Section \ref{Sec:Sys_model}. \textcolor{black}{The performance analysis for the perfect CSI case is elaborated in Section \ref{Sec:Achie}, while discussions on special and asymptotic cases are provided in Section \ref{Sec:discussions}. Section \ref{Sec:Ext_ICSI} shows the extension to the imperfect CSI case.} In Section \ref{Sec:Simu}, simulation results are provided, and finally we conclude the paper in Section \ref{Sec:Conc}. For a matrix $\mathbf{A}$, \textcolor{black}{the} trace, transpose, Hermitian, conjugate, and inverse are \textcolor{black}{operators} denoted as $\mathrm{tr}\{\mathbf{A}\}$, $\mathbf{A}^T$, $\mathbf{A}^H$, $\mathbf{A}^*$, and ${\mathbf{A}}^{-1}$, respectively. Also, $a_{ij}$ and $\mathbf{a}_{i}$ denote the $(i,j)$th entry	and the $i$th column of $\mathbf{A}$. For vector $\mathbf{a}$, $\norm{\mathbf{a}}$ denotes the 2-norm, and $\mathrm{diag}\{\mathbf{a}\}$ denotes a diagonal matrix with the elements of $\mathbf{a}$ as its diagonal entries. The $N \times N$ identity matrix reads as $\mathbf{I}_N$. Notation $\mathbb{E}\{\cdot\}$ is the expectation operator. \textcolor{black}{For a complex variable, $\Re\{.\}$ and $\Im\{.\}$ denote the real and imaginary parts, respectively.} Finally, $\mathrm{mod}_{N}(x)$ denotes $x$  modulo $N$.

        \section{System Model} \label{Sec:Sys_model}

This work considers a MWRN consisting of $K$ single-antenna users which exchange their information via a multi-antenna relay with $N$ antennas where $N \gg 1$  and $N\ge K$. \textcolor{black}{Frequency-flat narrowband channels are assumed.} Let $\mathbf{H}=\mathbf{\tilde{H}}\mathbf{D}^\frac{1}{2}$ be the $N \times K$ channel matrix between the users and the relay, where $\mathbf{\tilde{H}}\in \mathbb{C}^ {N \times K}$ is the fast fading channel matrix whose entries are independently and identically
distributed (i.i.d.) circularly symmetric complex Gaussian  with zero-mean and unit-variance, i.e., $\mathcal{CN}(0,1)$ and $\mathbf{D}\in \mathbb{R}^{K\times K}$ is a diagonal matrix whose $k$th diagonal element denoted as $\beta_{k}$
stands for the large-scale fading of the channels from \textcolor{black}{user} $k$ to the relay. \textcolor{black}{We define} 
\begin{align}
\beta_{\rm sum}\triangleq {\rm tr} \{{\bf D}\}
=\sum_{k=1}^K\beta_k,\quad 
\beta_{\backslash i}=\sum_{k=1,k\ne i}^K\beta_k. \notag
\end{align}
Further, denote the $k$th columns of $\mathbf{\tilde{H}}$ and $\mathbf{H}$ as  $\mathbf{\tilde{h}}_{k}$ and $\mathbf{h}_{k}$ which are the fast fading and overall channel vectors from \textcolor{black}{user} $k$ to the relay, respectively. It is assumed that the relay has perfect CSI. \textcolor{black}{The imperfect CSI case is considered in Section \ref{Sec:discussions}.}

\textcolor{black}{In a MWRN, each user detects signals from all other $K-1$ users. Under the half-duplex mode, the} communications are composed of two phases: the multiple access (MAC) phase consisting of one time slot, and the broadcast (BC) phase consisting of $K-1$ time slots to enable each user to receive information from all others.  
\subsection{The MAC Phase and the ADC Quantization}
 In the MAC phase, all users transmit their information signals simultaneously to the relay. Denote the vector of \textcolor{black}{normalized} information symbols of the users as $\mathbf{x}\in\mathbb{C}^{K \times 1}$\textcolor{black}{, and the average transmit power of each user as $p_u$. This implies that all users have the same transmit power.} The $k$th element of $\mathbf{x}$ corresponds to the signal of \textcolor{black}{user} $k$ normalized to have unit power {\color{black}$\mathbb{E}_{x_k \in\mathcal{S}_k}\{|x_k|^2\}=1$} for $k\in\{1, 2, \cdots, K\}${\color{black}, where $\mathcal{S}_k$ is the modulation set of user $k$}. \textcolor{black}{Hence}, the \textcolor{black}{baseband representation of the received discrete-time analog-valued signal\footnote{\textcolor{black}{This is referred to as ``analog signal" for short afterwards.}}} at the relay, denoted as $\mathbf{r}_{\mathrm{a}}\in \mathbb{C}^{N\times 1}$, can be written as
\begin{equation}
 \mathbf{r}_{\mathrm{a}}=\sqrt{p_u}\mathbf{H}\mathbf{x}+\mathbf{z}_\mathrm{R}=\sum_{i=1}^{K}{\sqrt{p_u}{\mathbf{h}_i}{x_i}}+\mathbf{z}_{\mathrm{R}}, \label{Eq:analog_sig}
\end{equation} 
where $\mathbf{z}_{\mathrm{R}}\sim \mathcal{CN}(\mathbf{0}_{N}, \mathbf{I}_{N})$ is the vector of additive white Gaussian noise (AWGN) at the relay. Denote the $n$th element of $\mathbf{r}_{\mathrm{a}}$ as $r_{n,\mathrm{a}}$.

 We assume that each relay antenna is equipped with a radio-frequency chain including a pair of low-resolution ADCs for the in-phase and quadrature components. We consider a generic mixed-ADC structure in which the ADC pairs of different antenna can have different resolutions. Denote the ADC resolution for the $n$th antenna as $b_n$ bits which is a positive integer value between $b_{\min}$ and $b_{\max}$. Let $\mathbf{b}=[b_1, \cdots ,b_N]$ which is the resolution profile of the relay antennas. The ADC quantization corresponding to the $n$th antenna can be characterized by a set of $2^{b_n}+1$ quantization thresholds $\mathcal{T}_{b_n}=\{\tau_{n,0}, \tau_{n,1},\cdots, \tau_{n,2^{b_n}}\}$, where $-\infty=\tau_{n,0} < \tau_{n,1}< \cdots< \tau_{n,2^{b_n}}=\infty$, and a set of $2^{b_n}$ quantization labels $\mathcal{L}_{b_n}=\{l_{n,0}, l_{n,1}, \cdots, l_{n,2^{b_n}-1}\}$, where $l_{n,i}\in (\tau_{n,i},\tau_{n,i+1}]$. 
  We describe the joint operation of the $n$th ADC pair at the relay by the function $\mathcal{Q}_{b_n}(\cdot):\mathbb{C}\rightarrow\mathcal R_{b_n}$, where $\mathcal{R}_{b_n}\triangleq\mathcal{L}_{b_n}\times\mathcal{L}_{b_n}$.   We denote the quantized signal vector at the relay by $\mathbf{\hat{r}}$  with $\hat{r}_n$ being its $n$th entry corresponding to the $n$th antenna. The quantization function $\mathcal{Q}_{b_n}(\cdot)$ maps the analog received signal, $r_{n,\mathrm{a}}$, to the quantized signal, $\hat{r}_n$, in a way that 
  \begin{align*}
  {\hat{r}_n}=\mathcal{Q}_{b_n}(r_{n,\mathrm{a}})=l_{n,k}+jl_{n,p}, \text{ if $\Re\{r_{n,\mathrm{a}}\}\in(\tau_{n,k},\tau_{n,k+1}]$ and $\Im\{r_{n,\mathrm{a}}\}\in (\tau_{n,p},\tau_{n,p+1}]$.}
  \end{align*} 
  Therefore, the quantized vector at the relay is 
 \begin{equation}
 \mathbf{\hat{r}}=\mathcal{Q}(\mathbf{r}_{\mathrm{a}})=\mathcal{Q}(\sqrt{p_u}\mathbf{H}\mathbf{x}+\mathbf{z}_\mathrm{R}),
 \end{equation}  
   where $\mathcal{Q}(\cdot)$ is the function that quantizes the $n$th entry of its input vector using $\mathcal{Q}_{b_n}(\cdot)$ for $n\in \{1,2,\cdots, N\}$.
   
The optimal sets of $\mathcal{L}_{b_n}$ and $\mathcal{T}_{b_n}$ for $n\in \{1, 2, \cdots, N\}$ that minimize the MSE between the non-quantized received vector $\mathbf{r}_{\mathrm{a}}$ and the quantized vector $\hat{\mathbf{r}}$ depends on the distribution of the input ${\mathbf{r}_{\mathrm{a}}}$, which changes with respect to the channels and the information signals. From a practical point of view \cite{jacobsson2017throughput}, we use the set of quantization labels and the set of thresholds that are optimal for Gaussian signals\textcolor{black}{\footnote{\textcolor{black}{The Gaussian assumption is accurate in the low-SNR regime or when the number of users is sufficiently large \cite{mezghani2012capacity}. Simulation results have verified the validity of this assumption for normal ranges of user number and SNR.}}}. From (\ref{Eq:analog_sig}), the variance of each entry of $\mathbf{r}_{\mathrm{a}}$ can be straightforwardly calculated to be 
\begin{equation}
v\triangleq 1+p_u\beta_{\mathrm{sum}}. \label{eq:v}
\end{equation} 
Then, using Lloyd-Max algorithm \cite{max1960quantizing,lloyd1982least}, we can find the \textcolor{black}{optimal} sets of labels and thresholds, \textcolor{black}{$\mathcal{L}_{b_n}^{*}=\{l_{n,0}^*, l_{n,1}^*, \cdots, l_{n,2^{b_n}-1}^*\}$} and \textcolor{black}{$\mathcal{T}_{b_n}^*=\{\tau_{n,0}^*, \tau_{n,1}^*,\cdots, \tau_{n,2^{b_n}}^*\}$}, respectively, that minimize the MSE when the analog signal follows $\mathcal{CN}(0,v)$. 

With the  set of labels \textcolor{black}{$\mathcal{L}_{b_n}^{*}$}, the set of thresholds \textcolor{black}{$\mathcal{T}_{b_n}^*$}, \textcolor{black}{and the Gaussian assumption of the quantization input}, the variance of the $n$th entry of $\mathbf{\hat{r}}$ with ADC resolution $b_n$, denoted as $C_{n,\hat{r}}$, can be straightforwardly obtained {\color{black}based on the definition of variance, quantization function, and Gaussian distribution of the input,} as 
\begin{align}
C_{n,\hat{r}}=\sum_{i=0}^{2^{b_n}-1}\textcolor{black}{{l_{n,i}^*}^2}\left[\left(\mathrm{erf}\left(\frac{\textcolor{black}{\tau_{n,i+1}^*}}{\sqrt{v}}\right)-\mathrm{erf}\left(\frac{\textcolor{black}{\tau_{n,i}^*}}{\sqrt{v}}\right)\right)\right], \label{Eq:quan_var} 
\end{align} 
where $\mathrm{erf}(\cdot)$ is the error function defined as $\mathrm{erf}(x)\triangleq\frac{1}{\sqrt{\pi}}\int_{-x}^{x} \mathrm{e}^{-t^2}dt$. Thus, the covariance matrix \textcolor{black}{and the average power of the received quantized vector at the relay are respectively,}
\begin{align}
\mathbf{C}_\mathbf{\hat{r}}=\mathrm{diag}\{C_{1,\hat{r}}, C_{2,\hat{r}}, \cdots, C_{N,\hat{r}}\}, \label{eq:hat_c_1}
\end{align}
\begin{eqnarray}
\hat{c}\triangleq\frac{1}{N}\mathrm{tr}\{\mathbf{C}_\mathbf{\hat{r}}\}
=\frac{1}{N}\sum_{n=1}^N C_{n,\hat{r}}. \label{eq:hat_c_2}
\end{eqnarray}
 \subsection{The BC Phase}
  The BC phase takes $K-1$ time slots.  {\color{black}ZF relay beam-forming \cite{amah2010beamforming} is used, in which the beam-forming matrix for the $t$th time slot is}
 \begin{align}
 \mathbf{G}^{(t)}&=\sqrt{\alpha^{(t)}}\mathbf{H}^*(\mathbf{H}^T\mathbf{H}^*)^{-1}\mathbf{P}^t(\mathbf{H}^H\mathbf{H})^{-1} \mathbf{H}^H,\label{equation:3.16}
 \end{align} 
where $\mathbf{P}$ is the permutation matrix obtained by shifting the columns of $\textbf{I}_K$ circularly to the right one time, and $\alpha^{(t)}$ is the \textcolor{black}{ZF} transmit power coefficient. {\color{black}Then, the transmit signal at the relay is}
 \begin{equation} 
 {\mathbf{r}^{(t)}_{\mathrm{t}}}=\mathbf{G}^{(t)}\mathbf{\hat{r}}. \label{equation:3.3}
 \end{equation} 
Let $P_{R}$ denote the average transmission power of the relay. {\color{black}Then, $\alpha^{(t)}$ must satisfy $P_R=\mathbb{E}\{\lVert{{\mathbf{r}^{(t)}_{\mathrm{t}}}}\rVert^2\}$.} With channel reciprocity, the channel from the relay to the users is $\mathbf{H}^T$. Thus, \textcolor{black}{the} received signal vector of all users in the BC time slot $t$, ${\mathbf{r}_{\mathrm{u}}^{(t)}}$, is
 \begin{align} \label{eq:r_u}
 {\mathbf{r}_{\mathrm{u}}^{(t)}}&=\mathbf{H}^T{\mathbf{r}^{(t)}_{\mathrm{t}}}+\mathbf{z}^{(t)}_{\mathrm{u}}, 
 \end{align}
 where $\mathbf{z}_{\mathrm{u}}^{(t)}=[{z_1}^{(t)}, {z_2}^{(t)},..., {z_K}^{(t)}]^T$ is the noise vector at the users whose elements are i.i.d. $\mathcal{CN}(0,1)$.
In the BC time slot $t$, \textcolor{black}{user} $k$ is supposed to decode \textcolor{black}{user} \textcolor{black}{$i(k,t)$}'s information symbol, where
\begin{align}
         \textcolor{black}{i(k,t)}\triangleq\mathrm{mod}_K(k+t-1)+1, \label{eq:i}
\end{align}
\textcolor{black}{which is a function of the receiving user's index, $k$, and the time slot, $t$. To help the presentation, it is simplified to $i(k)$ when there is no confusion.}

        \section{Average Achievable Rate \textcolor{black}{Analysis}} \label{Sec:Achie}
 {\color{black}This section considers the perfect CSI case. We first} analyze the ADC quantization process, then derive the relay power coefficient for the ZF beam-forming in each BC time slot, and finally, obtain the average achievable rate of the MWRN.

{\color{black}In general, the quantization at the low-resolution ADCs leads to signal distortion that is correlated with the quantization input signal. According to Bussgang's theorem \cite{bussgang1952crosscorrelation}, when the input to the ADCs is Gaussian, the quantized output can be written as a linear combination of the quantization input signal, $\mathbf{r}_{\mathrm{a}}$, and a quantization distortion, $\mathbf{d}$, that is uncorrelated to the quantization input signal. Thus,} the quantized received signal vector at the relay can be written as 
  \begin{align}
 \hat{r}_{n}&=\mathcal{Q}_{b_n}(r_{n,\mathrm{a}})\approx G_{b_n}r_{n,\mathrm{a}}+d_n, \text{or} \notag\\
 \mathbf{\hat{r}}&=\mathcal{Q}(\mathbf{r}_{\mathrm{a}})\approx\mathbf{G}_b\mathbf{r}_{\mathrm{a}}+\mathbf{d}, \label{eq:quantized_signal}
 \end{align} 
 where $G_{b_n}$ is the quantization coefficient corresponding to the $n$th ADC pair and  $\mathbf{G}_b=\\\mathrm{diag}\{G_{b_1}, \cdots, G_{b_N}\}$. For an arbitrary $n$, the value of $G_{b_n}$ can be calculated to be
 \begin{align}
 G_{b_n}=\frac{1}{\sqrt{\pi v}}\sum_{i=0}^{2^{b_n}-1}\textcolor{black}{l_{n,i}^*}\left[\exp\left(-\frac{\textcolor{black}{{\tau_{n,i}^*}^2}}{v}\right)-\exp\left(-\frac{\textcolor{black}{{\tau_{n,i+1}^*}^2}}{v}\right)\right]. \label{eq:buss}
 \end{align}
 \textcolor{black}{We define} 
\begin{align*}
&g_1\triangleq \frac{1}{N}\mathrm{tr}\{\mathbf{G}_b\}
=\frac{1}{N}\sum_{n=1}^NG_{b_n}, \text{and} \quad
g_2\triangleq \frac{1}{N}\mathrm{tr}\{\mathbf{G}_b^2\}
=\frac{1}{N}\sum_{n=1}^NG^2_{b_n},
\end{align*}
which are needed for the achievable rate expression. {\color{black}They represent the average of the quantization coefficients and quantization coefficients squared, respectively.}

The variance of each entry of $\mathbf{r}_{\mathrm{a}}$ is $v$, so the covariance matrix of the quantization distortion, $\mathbf{C}_\mathbf{d}=\mathbb{E}[\mathbf{d}\mathbf{d}^H]$, is
 \begin{align}
 \begin{split}
 \mathbf{C}_\mathbf{d}&\approx\mathbf{C}_\mathbf{\hat{r}}-\mathbb{E}\{\mathbf{G}_b\mathbf{r}_\mathrm{a}\mathbf{r}^H_\mathrm{a}\mathbf{G}_b\}=\mathbf{C}_\mathbf{\hat{r}}-v\mathbf{G}^2_b. \label{Eq:C_d}
 \end{split}
 \end{align}
  Next, we calculate the \textcolor{black}{ZF power coefficient}. The result is given in the following {\color{black}theorem}.

   \begin{Theo.}\label{lem:alpha}
        For a multiuser massive MIMO MWRN with $K$ single-antenna users, $N$ relay antennas, mixed-ADC with resolution profile $\mathbf{b}$ at the relay, relay power constraint $P_R$, and perfect CSI, the relay power coefficient for ZF beam-forming in time slot $t$ is 
\begin{align}
        	\alpha^{(t)}&\approx{{\frac{P_R(N-K)}{p_ug_1^2\sum_{m=1}^{K}\frac{1}{\beta_m} +\frac{NK+N-K^2}{(N-K)^2(K+1)}\left(\hat{c}-p_u\beta_{\rm sum}g_2\right)
        				\sum_{m=1}^{K}\frac{1}{\beta_m\beta_{i(m)}}}}}.\label{eq:p_zf_coeff}
\end{align}
  \end{Theo.}
\noindent  \textit{Proof:} Please see Appendix \ref{App:proof_p_zf}.
         
In what follows, the average achievable rate from \textcolor{black}{user} $i(k)$ to \textcolor{black}{user} $k$ will be derived where $i(k)$ is \textcolor{black}{given} in (\ref{eq:i}).

\noindent From (\ref{eq:r_u}) and Bussgang's decomposition in (\ref{eq:quantized_signal}), the received signal vector at the users in the BC time slot $t$ can be written as
\begin{align} \label{equation:3.6}
\begin{split}
{\mathbf{r}_{\mathrm{u}}^{(t)}}\hspace{-1mm}\approx \hspace{-1mm}\sqrt{p_u}\mathbf{H}^T\mathbf{G}^{(t)}\mathbf{G}_b\mathbf{H}\mathbf{x}
\hspace{-1mm}+\hspace{-1mm}\mathbf{H}^T\mathbf{G}^{(t)}\mathbf{G}_b\mathbf{z}_\mathrm{R}
\hspace{-1mm}+\hspace{-1mm}\mathbf{H}^T\mathbf{G}^{(t)}\mathbf{d}
\hspace{-1mm}+\hspace{-1mm}\mathbf{z}^{(t)}_{\mathrm{u}}.
\end{split}
\end{align}
 \textcolor{black}{Hence}, the received signal by \textcolor{black}{user} $k$, denoted as $ {r_{k}}^{(t)}$, is
 \begin{align}  
 \begin{split}
 {r_{k}}^{(t)} \approx&\sqrt{p_u}{\mathbf{h}_{k}}^{T}\mathbf{G}^{(t)}\mathbf{G}_b\textbf{h}_{i(k)}x_{i(k)}+\sqrt{p_u}{\mathbf{h}_{k}}^{T}\mathbf{G}^{(t)}\mathbf{G}_b\sum_{j=1, j\ne i(k) }^{N}{\mathbf{h}_jx_j}\\
 &+{\mathbf{h}_{k}}^{T}\mathbf{G}^{(t)}\mathbf{G}_b\mathbf{z}_{\mathrm{R}}+\mathbf{h}_{k}^T\mathbf{G}^{(t)}\mathbf{d}+ {z_k}^{(t)}, \label{equation:3.10}
 \end{split}
 \end{align}
  where in (\ref{equation:3.10}), the first, second, third, and fourth terms are the desired signal, the interference from other users, the noise propagated from the relay, and the quantization distortion propagated from the relay, respectively. Thus, the interference-plus-noise power is
   \begin{align}
   \begin{split}
      I_{k,i(k)}=&p_u\sum_{j=1, j\ne i(k) }^{K}{\abs{\mathbf{h}_{k}^{T}\mathbf{G}^{(t)}\mathbf{G}_b\mathbf{h}_j}^2}+\norm{{\mathbf{h}_{k}}^{T}\mathbf{G}^{(t)}\mathbf{G}_b}^2+\norm{{\mathbf{h}_{k}}^{T}\mathbf{G}^{(t)}\mathbf{d}}^2+1. \label{eq:interfer_noise}
   \end{split}
  \end{align}
  
It can be seen from the first term in (\ref{eq:interfer_noise}) that due to the mixed-ADC structure at the relay, the user interference is not fully eliminated by the {\color{black}ZF} design in (\ref{equation:3.16}). For the special case of uniform-ADC structure, we have $\mathbf{G}_b=G_b\mathbf{I}_N$, and the user interference can be fully eliminated. 
     
The average achievable rate from \textcolor{black}{user} $i(k)$ to \textcolor{black}{user} $k$, {\color{black}denoted by}  $R_{k,i(k)}$, is given as
  \begin{align}
  R_{k,i(k)}\approx\mathbb{E}
  \left\{\log_2\left(1+\dfrac{p_u \abs{{\mathbf{h}_{k}}^{T}\mathbf{G}^{(t)}\mathbf{G}_b\mathbf{h}_{i(k)}}^2}{I_{k,i(k)}}\right)\right\}. \label{eq:avg_rate}
  \end{align}

Next, the average achievable rate of two arbitrary users in the MWRN is derived and the result is presented in the following {\color{black}theorem}.

\begin{Theo.} \label{theo:Theorem_ach}
        For a multiuser massive MIMO MWRN with $K$ single-antenna users, $N$ relay antennas, ZF beam-forming in (\ref{equation:3.16}), mixed-ADC with resolution profile $\mathbf{b}$ at the relay, and perfect CSI, the average achievable rate from \textcolor{black}{user} $i(k)$ to \textcolor{black}{user} $k$ is 
\begin{align} 
R_{k,i(k)}&\approx\log_2\left(1+\frac{F_1}{F_2}\right),
\label{eq:Rate}
\end{align}
where
\begin{align}
\begin{split}
F_1\triangleq&p_u\beta_{i(k)}\left[\frac{(NK+N-K^2-2K)g_1^2+Kg_2}{K+1}\right],\\
F_2\triangleq &\hat{c}+\frac{(N-K)}{\alpha^{(t)}}\beta_{i(k)}-p_u\beta_{\backslash i(k)}g_1^2 -p_u\beta_{i(k)}g_2.
\end{split}
\label{def-F2}
\end{align}
\end{Theo.}
\noindent\textit{Proof:} Please see Appendix \ref{App:proof_rate}.

The formula in (\ref{eq:Rate}) and parameter values in (\ref{eq:p_zf_coeff}) and (\ref{def-F2}) show how quantitatively the average achievable rate is affected by system settings such as the user power, relay power, number of users, and number of relay antennas. The effect of the ADC resolution is shown via the parameters $g_1$ and $g_2$. Specifically, it can be seen that the average achievable rate increases as $p_u$ or $P_R$ increases. 
{\color{black}\section{Results for Asymptotic Cases and Uniform-ADC}\label{Sec:discussions}
}
In what follows, we discuss two asymptotic cases and the special uniform-ADC case to gain insights on the effects of other parameters on the average achievable rate. 
\subsection{Asymptotic Cases}\label{sec:discussion}
The first asymptotic case {\color{black}commonly considered in mMIMO} is when $N\rightarrow\infty$ with fixed $K$. The result {\color{black}on the achievable rate can be simplified from (\ref{eq:p_zf_coeff}) (\ref{eq:Rate}) (\ref{def-F2}) as} \begin{align}
&R_{k,i(k),1}\approx \log_2\left(1+\frac{Np_u\beta_{i(k)}g_1^2}
{\hat{c}+\frac{p_u}{P_R}\beta_{i(k)}\left(\sum_{m=1}^{K}\frac{1}{\beta_m}\right)
	g_1^2-p_u\beta_{\backslash i(k)}g_1^2-p_u\beta_{i(k)}g_2
}\right){\color{black}.}\label{eq:largeN}
\end{align} {\color{black}It} shows that the signal-to-interference-plus-noise ratio (SINR) increases linearly in $N$. {\color{black}The effect of ADC-resolution is through $\hat{c}$, $g_1^2$, and $g_2$. Further, the expression directly shows that the sum-rate is monotonically increasing with respect to $p_u$ and $p_R$, but with finite ceiling as $p_u$ $\rightarrow$ $\infty$ or $p_R$ $\rightarrow$ $\infty$.  The ceiling depends on the ADC resolution profile, indicating that the penalty brought by low-resolution ADCs may not be fully compensated by increasing the user or relay transmit power.}

The second asymptotic case is when $N,K\rightarrow\infty$ with fixed ratio $K/N=c$. The constant $c$ is referred to as the loading factor. Notice that $\beta_{\rm sum}$ is linear in $K$, thus $v$ given in (\ref{eq:v}) and $\hat{c}$ given in (\ref{eq:hat_c_1}) and (\ref{eq:hat_c_2}) are also linear in $K$. We assume that as $K\rightarrow \infty$, the values $\bar{\hat{c}}\triangleq{\hat{c}}/{K}$, $\bar{\beta}\triangleq{\beta_{\mathrm{sum}}}/{K}$, $\bar{\beta}_{-1}\triangleq\sum_{m=1}^K{1}/{\beta_m}$, and $\bar{\beta}_{-2}\triangleq\sum_{m=1}^K{1}/({\beta_m\beta_{i(m)}})$ all converge to positive constants. It can be shown that $\lim_{K\rightarrow \infty}\left(\bar{\beta}-{\beta_{\backslash i(k)}}/{K}\right)=0$. Thus, for this case, {\color{black}the ZF power coefficient is} 
\[
\alpha^{(t)}_2\approx {\frac{NP_R}{\frac{1}{1-c}p_ug_1^2\bar{\beta}_{-1} 
		+\frac{c}{(1-c)^2}\left(\bar{\hat{c}}-p_u\bar{\beta} g_2\right)
		\bar{\beta}_{-2}}},\] 
and the average achievable rate is simplified as
\begin{align}
&R_{k,i(k),2}\approx \log_2\left(1+\frac{(1-c)}{c}\cdot\frac{p_u\beta_{i(k)}g_1^2}
{(\bar{\hat{c}}-p_u\bar{\beta} g_1^2)}\right).\label{eq:largeN_prop_K} 
\end{align}
 It shows that when the number of users increases linearly with the number of relay antennas, the average achievable rate becomes independent of $N$ and decreases as $c$ increases. The SINR for this asymptotic case is linear in $(1-c)/c$. {\color{black} The effect of ADC resolution profile is shown through $\bar{\hat{c}}$ and $g_1^2$. The expression indicates that the achievable rate degradation caused by low-resolution ADCs can be compensated by decreasing $c$. It also shows that the sum-rate is monotonically increasing with respect to $p_u$, but with a finite ceiling as $p_u\rightarrow$ $\infty$. The ceiling depends on the ADC resolution profile, indicating that the penalty brought by low-resolution ADCs may not be fully compensated by increasing the user power. Finally, it indicates that the average achievable rate in this case is independent of $p_R$.}

\subsection{Achievable Rate for the Uniform-ADC Case} \label{sub:uniform}
 {\color{black}For} the {\color{black}special case of} uniform-ADC with $b$-bit {\color{black}resolution}, denote the quantization coefficient as $G_b$ and the corresponding variance of the quantization output as $C_{\hat{r}}$. We have $G_{b_n}=G_b$ and $C_{n,\hat{r}}=C_{\hat{r}}$ for all $n$. Consequently, the {\color{black}ZF} relay power coefficient and the average achievable rate can be simplified as 
\begin{align}
&\alpha^{(t)}_{\mathrm{uni}}\approx{{\frac{P_R(N-K)}{\mathlarger\sum_{m=1}^{K}\frac{1}{\beta_m}
			\left[ p_u{G}_b^2+\frac{1}{\beta_{i(m)}}(C_{\hat{r}}-p_u\beta_{\mathrm{sum}}G^2_{b})\frac{(NK+N-K^2)}{(N-K)^2(K+1)}\right]}}},\label{eq:p_zf_coeff_uni} \\
&R_{k,i(k),\mathrm{uni}}\approx\log_2\left(1+\frac{(N-K)\beta_{i(k)}}{(\sqrt{p_u}{G}_b)^{-2}
	\left[\frac{N-K}{\alpha^{(t)}_{\mathrm{uni}}}\beta_{i(k)}+C_{\hat{r}}\right]
	-\beta_{\mathrm{sum}}}\right).\label{eq:Rate_uni}
\end{align}
The result reveals that the relay power, the quantization coefficient squared, and the user power have similar effect on the achievable rate. Increasing each of them has a positive effect on the achievable rate with a negative acceleration. Furthermore, $G_b^2$ and $p_u$ appear together as a product in the formulas. This means that they can be adjusted to compensate each other's contribution to the achievable rate. Another important fact that can be concluded from (\ref{eq:p_zf_coeff_uni}) and (\ref{eq:Rate_uni}) is that the achievable rate linearly decreases with the number of users while it has an increasing relation with the number of antennas.  
\textcolor{black}{\section{Extension to the Imperfect CSI Case}\label{Sec:Ext_ICSI}}
\textcolor{black}{In this section, we extend our results to the imperfect CSI case with the following widely used channel model\textcolor{black}{\footnote{\color{black}Channel estimation schemes for MIMO systems with low- or mixed-resolution ADCs can be found in \cite{jacobsson2015one,  liang2016mixed,choi2016near,mo2018channel}.}}: $\mathbf{\tilde{H}}=\mathbf{\hat{\tilde{H}}}+\Delta\mathbf{\tilde{H}}$, where $\mathbf{\hat{\tilde{H}}}\sim\mathcal{CN}(\mathbf{0},(1-\sigma_e^2)\mathbf{I}_N)$ is the estimation of the  small-scale fading channel, $\Delta\mathbf{\tilde{H}}\sim\mathcal{CN}(\mathbf{0},\sigma_e^2\mathbf{I}_N)$ is the CSI error, and $\sigma_e^2$ represents the power of the CSI error. Furthermore, $\mathbf{\hat{\tilde{H}}}$ and $\Delta\mathbf{\tilde{H}}$ are assumed to be independent. ZF relay beam-forming matrix considering the channel estimation $\mathbf{\hat{\tilde{H}}}$, is
 \begin{align}
	\mathbf{\hat{G}}^{(t)}&=\sqrt{\hat{{\alpha}}^{(t)}}\mathbf{\hat{H}}^*(\mathbf{\hat{H}}^T\mathbf{\hat{H}}^*)^{-1}\mathbf{P}^t(\mathbf{\hat{H}}^H\mathbf{\hat{H}})^{-1} \mathbf{\hat{H}}^H,\label{eq:ICSI_ZF}
	\end{align}
\noindent where $\mathbf{\hat{H}}=\mathbf{\hat{\tilde{H}}}\mathbf{D}^{\frac{1}{2}}$.} \textcolor{black}{ Therefore, the received signal by user $k$ is
	\begin{align}  
	\begin{split}
	{r_{k,\mathrm{ICSI}}}^{(t)} \approx&\sqrt{p_u}{\mathbf{h}_{k}}^{T}\mathbf{\hat{G}}^{(t)}\mathbf{G}_b\textbf{h}_{i(k)}x_{i(k)}+\sqrt{p_u}{\mathbf{h}_{k}}^{T}\mathbf{\hat{G}}^{(t)}\mathbf{G}_b\sum_{j=1, j\ne i(k) }^{N}{\mathbf{h}_jx_j}\\
	&+{\mathbf{h}_{k}}^{T}\mathbf{\hat{G}}^{(t)}\mathbf{G}_b\mathbf{z}_{\mathrm{R}}+\mathbf{h}_{k}^T\mathbf{\hat{G}}^{(t)}\mathbf{d}+ {z_k}^{(t)}, \label{eq:trans_ICSI}
	\end{split}
	\end{align} and the power of the interference-plus-noise terms is
	\begin{align}
	\begin{split}
	{I}_{k,i(k),\mathrm{ICSI}}=&p_u\sum_{j=1, j\ne i(k) }^{K}{\abs{\mathbf{h}_{k}^{T}\mathbf{\hat{G}}^{(t)}\mathbf{G}_b\mathbf{h}_j}^2}+\norm{{\mathbf{h}_{k}}^{T}\mathbf{\hat{G}}^{(t)}\mathbf{G}_b}^2+\norm{{\mathbf{h}_{k}}^{T}\mathbf{\hat{G}}^{(t)}\mathbf{d}}^2+1. \label{eq:interfer_noise_ICSI}
	\end{split}
	\end{align} Therefore, the average achievable rate from user $i(k)$ to user $k$ is
	\begin{align}
	{R}_{k,i(k),\mathrm{ICSI}}\approx\mathbb{E}
	\left\{\log_2\left(1+\dfrac{p_u \abs{{\mathbf{h}_{k}}^{T}\mathbf{\hat{G}}^{(t)}\mathbf{G}_b\mathbf{h}_{i(k)}}^2}{I_{k,i(k),\mathrm{ICSI}}}\right)\right\}. \label{eq:avg_rate_ICSI}
	\end{align}} \textcolor{black}{By following similar steps as in Appendix \ref{App:proof_p_zf} and modifying the SVD in (\ref{eq:sing_val_dec}) as  $\mathbf{\hat{\tilde{H}}}=\mathbf{U}\mathbf{\hat{\Sigma}\mathbf{V}^H}$, where $\mathbf{\hat{\Sigma}}=\sqrt{1-\sigma_e^2}\mathbf{\Sigma}$, the relay ZF transmit power coefficient for the imperfect CSI case is found as in Theorem \ref{lem:alpha_ICSI}.
 \begin{Theo.}\label{lem:alpha_ICSI}
 	For a multiuser massive MIMO MWRN with $K$ single-antenna users, $N$ relay antennas, mixed-ADC with resolution profile $\mathbf{b}$ at the relay, relay power constraint $P_R$, and imperfect CSI, the relay ZF power coefficient in the BC time slot $t$ is 
 	\begin{align}
 	\hat{{\alpha}}^{(t)}&\approx{{\frac{P_R(N-K)(1-\sigma_e^2)}{p_ug_1^2\sum_{m=1}^{K}\frac{1}{\beta_m} +\frac{NK+N-K^2}{(N-K)^2(K+1)(1-\sigma_e^2)}\left(\hat{c}-p_u\beta_{\rm{sum}}g_2\left(1-\sigma_e^2\right)\right)
 				\sum_{m=1}^{K}\frac{1}{\beta_m\beta_{i(m)}}}}}.\label{eq:p_zf_coeff_ICSI}
 	\end{align}
 \end{Theo.}
\noindent Further, by following similar steps as in Appendix \ref{App:proof_rate}, the average achievable rate can be found as in Theorem \ref{theo:Theorem_ach_ICSI}.
\begin{Theo.} \label{theo:Theorem_ach_ICSI}
	For a multiuser massive MIMO MWRN with $K$ single-antenna users, $N$ relay antennas, mixed-ADC with resolution profile $\mathbf{b}$ at the relay, imperfect CSI, and ZF beam-forming in (\ref{eq:ICSI_ZF}), the average achievable rate from \textcolor{black}{user} $i(k)$ to \textcolor{black}{user} $k$ is 
	\begin{align} 
	R_{k,i(k)}&\approx\log_2\left(1+\frac{\hat{F}_1}{\hat{F}_2}\right),
	\label{eq:Rate_ICSI}
	\end{align}
	where
{\color{black}	
\begin{align}
	\begin{split}
\hat{F}_1\triangleq& \frac{p_u\beta_{i(k)}}{K+1}\left[\vphantom{\sum_{m=1}^{K}\frac{1}{\beta_m\beta_{i(m)}}} g_1^2(1-\sigma_e^2)^2(NK+N-K^2-2K)+g_2\left(\vphantom{\sum_{m=1}^{K}\frac{1}{\beta_m\beta_{i(m)}}} (1-\sigma_e^2)(K+\sigma_e^2)+\right.\right.\\
	&\left.\left.\frac{(NK+N-K^2)}{(N-K)^2}\beta_k\beta_{i(k)}\sigma_e^4\sum_{m=1}^{K}\frac{1}{\beta_m\beta_{i(m)}}\right)\right],\\
	\hat{F}_2\triangleq& \hat{c}+\frac{(N-K)(1-\sigma_e^2)}{\hat{{\alpha}}^{(t)}}\beta_{i(k)}-p_u\beta_{i(k)}g_2+p_ug_1^2\left(\vphantom{\sum_{m=1}^{K}\frac{1}{\beta_m\beta_{i(m)}}}-\beta_{\backslash i(k)}(1-\sigma_e^2)+\right.\\ &\left. \beta_{i(k)}\beta_k\sigma_e^2\sum_{j=1,j\ne k}^{K}\frac{1}{\beta_j} +\frac{\beta_{i(k)}\beta_k^2\sigma_e^4 N}{(N-K)K(1-\sigma_e^2)}\sum_{m=1}^{K}\frac{1}{\beta_m\beta_{i(m)}}\right).
	\label{def-F2_ICSI}
	\end{split}
	\end{align}}
\end{Theo.}}
{\color{black}As it can be seen from $\hat{F}_1$ and $\hat{F}_2$, the dominant terms in the numerator are the first and the second terms, while the dominant terms in the denominator are the first four terms with the first one being the most dominant. This means that the effect of decrease in the average achievable rate due to the channel estimation error, gets scaled by $g_1^2$ and $g_2$ in the numerator. In other words, the higher the average of the resolution profile, the higher the decrease in the average achievable rate due to the channel estimation error.}

        {\color{black}\section{Simulation Results}
	\label{Sec:Simu}}
This section shows simulation results on the average achievable rates for \textcolor{black}{mMIMO MWRNs}. {\color{black}The closed-form results in Theorems \ref{theo:Theorem_ach} and \ref{theo:Theorem_ach_ICSI} are compared with the Monte-Carlo simulated ones. Also, the asymptotic results in (\ref{eq:largeN})-(\ref{eq:largeN_prop_K})} are compared with the general theoretical results. While $R_{1,2}$, the average \textcolor{black}{achievable rate} of \textcolor{black}{user} $2$ at \textcolor{black}{user} $1$ is used, similar results can be obtained for other user pairs.

The simulation contains two parts: the quantizer optimization and the achievable rate simulation. {\color{black}A training set of $10^5$ points is generated based on complex Gaussian channels and quadrature amplitude modulation (QAM). Then,} Lloyd-Max algorithm is used to find the quantizer for each value in the resolution profile $\mathbf{b}$. {\color{black}For the second part, the quantizers obtained in the previous step are used. $10^3$ channel realizations are generated, and for each channel $10^2$ information vectors are generated. Unless otherwise mentioned, networks with homogeneous channels are considered, i.e., the same large-scale fading for all users, where we set $\beta_{k}=0$ dB for $k \in \{1, 2, \cdots ,K\}$.} 
\begin{figure}[t]
        \centering
        \includegraphics[width=12cm, height=9cm]{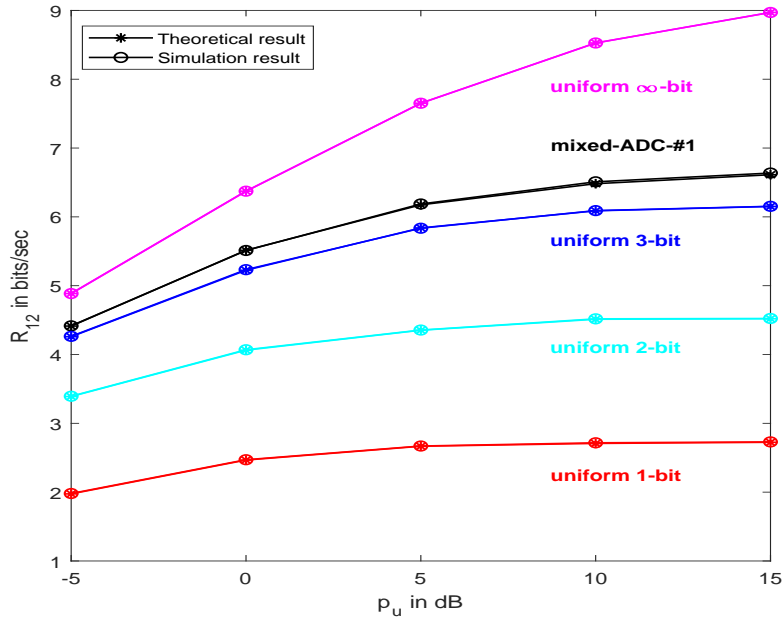}
        \vspace{-.5cm}
        \caption{Theoretical and simulation rate results versus average user power for different
         ADC profiles, $N=100$, $K=5$, and $P_R=15$ dB.}
        \label{fig:pu_q_jour}
        \vspace{-.5cm}
\end{figure}

\begin{figure}[t]
        \centering
        \includegraphics[width=12cm, height=9cm]{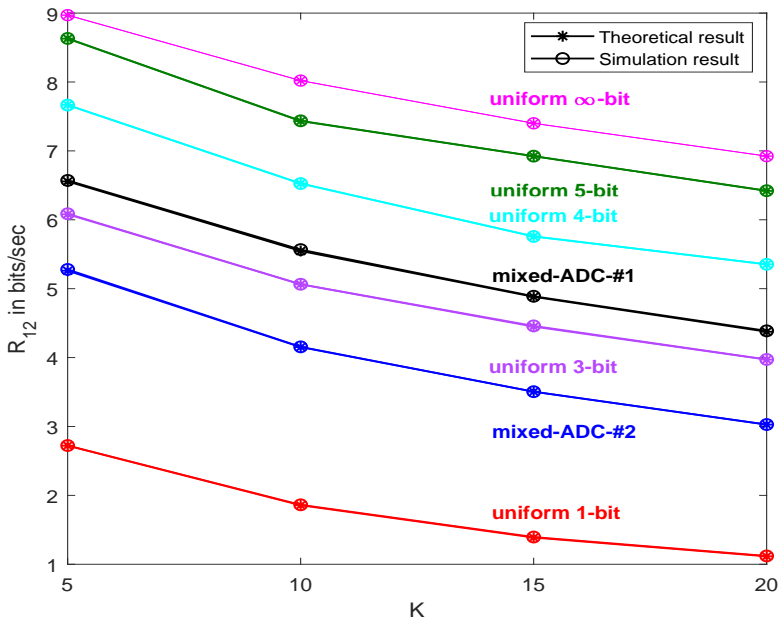}
        \vspace{-.5cm}
        \caption{Theoretical and simulation rate results versus the number of users for different ADC profiles, $N=100$, $p_u=P_R=15$ dB.} 
        \label{fig:K_q_jour}
        \vspace{-.5cm}
\end{figure}     
\begin{table}[h]
	\caption{Number of antennas with each resolution level in mixed-ADC-\#1 profile} \label{tab:q=8}
	\centering
	\begin{tabular}{|c|c|c|c|c|c|c|c|c|}
		\hline
		\backslashbox{N}{bits}  &1 & 2 & 3 & 4& 5& 6 & 7 & 8 \\
		\hline
		50&     3 & 6 & 7  & 6 & 7& 11& 5 & 5   \\
		\hline
		100 &   7 & 12 & 15  & 11 & 14& 22& 9 & 10   \\
		\hline
	\end{tabular}
\end{table}
\begin{figure}[h]
        \centering
        \includegraphics[width=12cm, height=9cm]{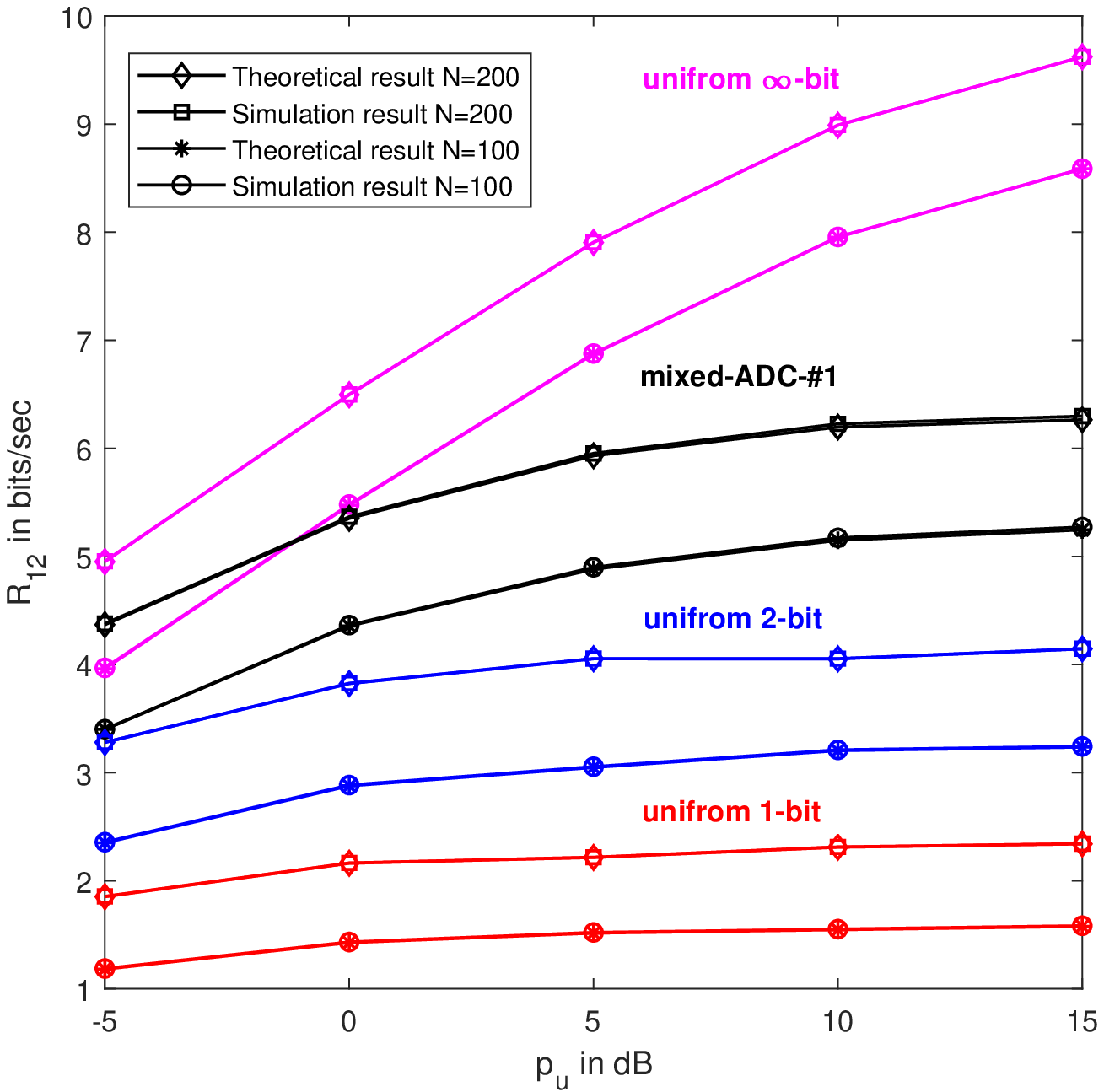}
        \vspace{-.5cm}
        \caption{Theoretical and simulation rate results versus average user power for \textcolor{black}{networks with heterogeneous channels} when $K=5$, $\beta_1=1, \beta_2=0.5, \beta_3=0.5, \beta_4=2, \beta_5=3$ and $P_R=15$ dB for $N=100$ and $N=200$.} 
        \label{fig:pu_q_heter_jour}
\end{figure}
In Figure \ref{fig:pu_q_jour}, a MWRN where $N=100$, $K=5$, and $P_R=15$ dB is considered where the simulation and theoretical results obtained by (\ref{eq:p_zf_coeff}) and (\ref{eq:Rate}) are compared when $p_u$ changes from $-5$ to 15 dB. Five resolution profiles are considered: uniform 1-bit, 2-bit, 3-bit, $\infty$-bit ADCs, and a mixed-ADC (referred to as mixed-ADC-\#1) specified in Table~\ref{tab:q=8}. {\color{black}In mixed-ADC-\#1 profile, the ADC resolution for each antenna is randomly and independently generated according to the discrete uniform distribution on $[1, 8]$. It is shown that the simulation and theoretical results perfectly match for all power range and ADC profiles. Also, the figure shows the rate degradation due to low-resolution ADCs, especially in the high SNR region. For instance, when $p_u=15$ dB, the achievable rate for mixed-ADC-\#1 is about $74\%$ of the full precision case, implying the importance of the ADC resolutions on the rate performance for {\color{black}mMIMO} MWRNs.} {\color{black} Finally, it can be observed that for all low-resolution ADC cases, as $p_u$ increases, the achievable rate saturates quickly.}

{\color{black}Figure \ref{fig:K_q_jour} shows the rate results when  $N=100$, $p_u=P_R=15$ dB, and  $K=5, 10, 15, 20$. Five uniform-ADC profiles with the bit levels of $1,3,4,5, \infty$ are tested. Also, two mixed-ADC profiles are examined: the mixed-ADC-\#1 explained in Table \ref{tab:q=8} and the mixed-ADC-\#2 for which the resolutions are 1 to 4 bits and the numbers of antennas are 21, 27, 22, and 30, respectively, for the 4 resolution levels. This figure confirms the perfect match between the simulation and theoretical results for all numbers of users and ADC profiles. It also reveals the degradation in the achievable rate with the increase in the number of users. This is due to the decrease in ZF beam-forming power scaling factor $\alpha^{(t)}$ that causes loss in the SINR. Further, Figures 1 and 2 show that the higher the average ADC resolution, the higher the average achievable rate.  

Next, a MWRN \textcolor{black}{with $5$ users and heterogeneous channels is considered where $\beta_1=1, \beta_2=0.5, \beta_3=0.5, \beta_4=2, \beta_5=3$, and $P_R=15$ dB.} Resolution profiles of uniform 1-bit, 2-bit, and $\infty$-bit along with the mixed-ADC-\#1 are considered. {\color{black}For $N=100$, the mixed-ADC-\#1 profile is shown in Table \ref{tab:q=8}.} For $N=200$ the number of ADC pairs for each resolution level is twice the number for $N=100$. {\color{black} Figure \ref{fig:pu_q_heter_jour} approves the prefect match between the derivations and simulation results for both cases when $p_u$ changes from $-5$ to $15$ dB.} {\color{black}An important observation here is that in the medium to high SNR region, increasing the ADC resolutions has higher impact on the rate compared to increasing the number of antennas. In other words, a large number of low-resolution ADCs cannot keep up with lower number of high-resolution ADCs in the sense of achievable rate.} 

\begin{figure}[t]
	\centering
	\includegraphics[width=12cm, height=9cm]{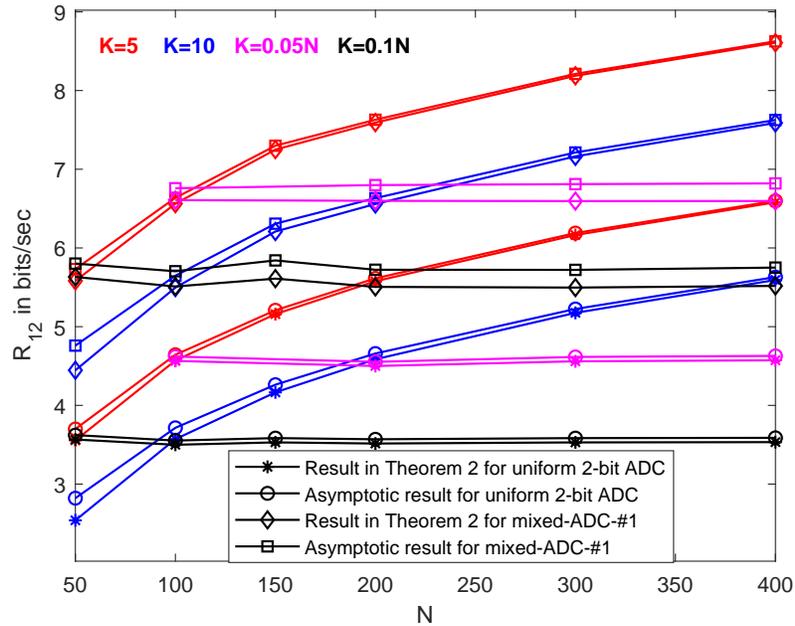}
	\vspace{-.5cm}
	\caption{Results in {\color{black}Theorem} \ref{theo:Theorem_ach} are compared with the asymptotic results in (\ref{eq:largeN}) when $K=5,10$ {\color{black}(in red and blue)}, and the asymptotic results in (\ref{eq:largeN_prop_K}) when $c=0.05, 0.1$ {\color{black}(in pink and black)}. In all simulations $p_u=P_R=15$ dB.} 
	\label{fig:K_N_q_asym_jour}
	\vspace{-.5cm}
\end{figure}
\vspace{-1pt}
\begin{figure}[h!]
	\centering
	\includegraphics[width=12cm, height=9cm]{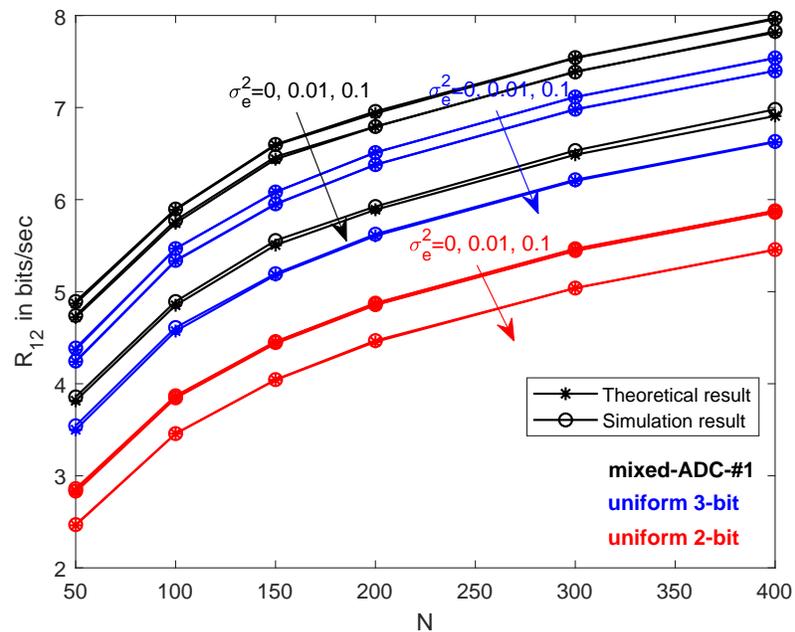}
	\vspace{-.5cm}
	\caption{\color{black}Theoretical and simulation rate results versus the number of relay antennas for the imperfect CSI cases, $\sigma_e^2=0, 0.01, 0.1$, $K=8$ and $p_u=P_R=15$ dB.} 
	\label{fig:K_N_q_ICSI}
	\vspace{-.5cm}
\end{figure} 
\vspace{-1pt}
{\color{black} Figure \ref{fig:K_N_q_asym_jour} compares the results in Theorem \ref{theo:Theorem_ach} with the two asymptotic results in (\ref{eq:largeN}) and (\ref{eq:largeN_prop_K}). For the first asymptotic case, $K=5, 10$ are tested and for the second one, $c=0.05, 0.1$ are tested. In all simulations, the number of antennas changes from $50$ to $400$ except for the $K=0.05N$ case that $N$ takes multiples of $100$. The results are shown for two ADC profiles: uniform $2$-bit and mixed-ADC-\#1. For mixed-ADC-\#1, the ADC profiles when $N=50, 100$ are specified in Table \ref{tab:q=8}. {\color{black}For $N=150, 200, 300, 400$ the ADC profiles are found by scaling the ADC profile for $N=50$  three times and the ADC profile for $N=100$, two, three, and four times, respectively. Figure \ref{fig:K_N_q_asym_jour} confirms that our asymptotic analysis for case 1 perfectly matches the general results for $N\ge200$, while it is a tight upper bound for $N< 200$. Also, the gap between the results from (\ref{eq:largeN}) and Theorem \ref{theo:Theorem_ach} shrinks as the number of users decreases.} In addition, Figure \ref{fig:K_N_q_asym_jour} indicates that the asymptotic result for case 2 works well for the uniform-ADC cases while for mixed-ADC cases, it is an upper bound with a small gap that shrinks as $c$ decreases. This figure also confirms that for asymptotic case 1, the rate linearly increases with $N$, while for case 2 it only increases if $c$ decreases. 
}  

{\color{black}Finally, for the imperfect CSI case the rate results versus the number of relay antennas are illustrated in Figure \ref{fig:K_N_q_ICSI} where $\sigma_e^2=0, 0.01, 0.1$, $K=8$, and $p_u=P_R=15$ dB. This figure shows that our result in (\ref{eq:p_zf_coeff_ICSI})-(\ref{def-F2_ICSI}) is accurate. Also, it shows that higher estimation error leads to lower rate. In addition, the higher the average resolution of the ADCs, the larger the gap between the rates of the perfect and imperfect CSI cases implying that higher resolution ADCs are more sensitive to CSI error.}

        \section{Conclusion}
\label{Sec:Conc} 
{\color{black}  In this paper, we have examined the multi-level mixed-ADC receive architecture in mMIMO MWRNs and derived tight closed-form approximations for the average achievable rates under ZF relay beam-forming considering both perfect and imperfect CSI. We have developed a new analytical method using SVD for Gaussian matrices, distributions of the singular values of Gaussian matrices, and properties of Haar matrices. The results characterize the achievable rate in terms of the system parameters and specifically, quantify the performance degradation caused by low-resolution ADCs and channel estimation error. It is shown that in the medium to high SNR region the ADC resolution has more significant effect on the rate compared to the number of antennas. Insightful asymptotic expressions are derived when the number of antennas grows towards infinity. Also, the special case of uniform-ADC is studied for comparison with the mixed-ADC case. Monte-Carlo simulations have verified the accuracy of our theoretical derivations. Our simulations show that the system with higher resolution ADCs is more sensitive to the CSI quality. Therefore, under channel estimation error, low-resolution ADCs are able to both keep the rate performance and save significant hardware cost and energy.}
 
        \appendices
\section{Proof of {\color{black}{Theorem}} \ref{lem:alpha}} \label{App:proof_p_zf}
{\color{black}The coefficient $\alpha^{(t)}$ is the solution of $P_R=\mathbb{E}\{\lVert{{\mathbf{r}^{(t)}_{\mathrm{t}}}}\rVert^2\}$, where the right-hand-side  can be written as the sum of several terms. In calculating each term, we first simplify the expression by the SVD of the channel matrix. Then, we use properties of Wishart distributed matrices, Haar distributed matrices, and the singular values for Gaussian matrix to calculate the value. The following lemma is provided on the the properties of Haar distributed matrices.}
\vspace{-.35cm}
\begin{Lemma}\label{lem:moments}
        If $1\le i,j,i',j'\le N$, $i \ne i', j \ne j'$, and $\mathbf{U}$ is an $N\times N$ Haar (isotropically) distributed matrix, then the following hold \cite{hiai1999asymptotic}.
{\color{black}   \begin{align}
		1) \hspace{0.1 cm} & \mathbb{E}(\lvert u_{i,j}\rvert^2)=\frac{1}{N},\hspace{0.5 cm}
		2) \hspace{0.1 cm} \mathbb{E}(\lvert u_{i,j}\rvert^4)=\frac{2}{N(N+1)},\notag\\
		3)\hspace{0.1 cm} & \mathbb{E}(\lvert u_{i,j}\rvert^2\lvert u_{i',j}\rvert^2)=\mathbb{E}(\lvert u_{i,j}\rvert^2\lvert u_{i,j'}\rvert^2)=\frac{1}{N(N+1)},\notag\\
		4)\hspace{0.1 cm}& \mathbb{E}(\lvert u_{i,j}\rvert^2\lvert u_{i',j'}\rvert^2)=\frac{1}{N^2-1},\hspace{0.5 cm}
		5)\hspace{0.1 cm}  \mathbb{E}( u_{i,j} u_{i',j'} u_{i,j'}^*u_{i',j}^*)=-\frac{1}{N(N^2-1)}.\notag
\end{align}}
All other multiple moments up to the fourth order are zero.
\end{Lemma}
\vspace{-.45cm}
 \noindent From (\ref{Eq:analog_sig}), (\ref{equation:3.3}), and (\ref{eq:quantized_signal}) we can write
\begin{align}
{\mathbf{r}^{(t)}_{\mathrm{t}}}&\approx\mathbf{G}^{(t)}\mathbf{G}_b(\sqrt{p_u}\mathbf{H}\mathbf{x}+\mathbf{z}_\mathrm{R})+\mathbf{G}^{(t)}\mathbf{d}.\notag
\end{align}
As $\mathbf{r}_{\mathrm{a}}$ and $\mathbf{d}$ are uncorrelated, we can write $\mathbb{E}\{\lVert{{\mathbf{r}^{(t)}_{\mathrm{t}}}}\rVert^2\}\approx c_1+c_2+c_3$ where
\begin{align}
c_1&\triangleq p_u\mathbb{E}[\mathrm{tr}\{\mathbf{G}^{(t)}\mathbf{G}_b\mathbf{H}\mathbf{H}^H\mathbf{G}_b(\mathbf{G}^{(t)})^H\}]\notag,\\ 
c_2&\triangleq\mathbb{E}[\mathrm{tr}\{\mathbf{G}^{(t)}\mathbf{G}_b^2(\mathbf{G}^{(t)})^H\}]\notag,\hspace{0.1 cm}
c_3\triangleq\mathbb{E}[\mathrm{tr}\{\mathbf{G}^{(t)}\mathbf{d}\mathbf{d}^H(\mathbf{G}^{(t)})^H\}].\notag
\end{align}
For the calculation of $c_1$, we use the following approximation}{\color{black}\footnote{\color{black} This approximation is obtained by replacing the $\mathbf{G}_b$ matrices on the left-hand side by $\frac{1}{N}\sum_{n=1}^{N} G_{b_n} \mathbf{I}_N$. Our simulation results show that this approximation is tight.}}
\begin{align}
        \mathbb{E}[\mathrm{tr}\{\mathbf{G}^{(t)}\mathbf{G}_b\mathbf{H}\mathbf{H}^H\mathbf{G}_b(\mathbf{G}^{(t)})^H\}\approx\notag \left(\frac{1}{N}\sum_{n=1}^{N}G_{b_{n}}\right)^2\mathbb{E}[\mathrm{tr}\{\mathbf{G}^{(t)}\mathbf{H}\mathbf{H}^H(\mathbf{G}^{(t)})^H\}]\notag.
\end{align}By using the $\mathbf{G}^{(t)}$ expression in (\ref{equation:3.16}), $c_1\approx \alpha^{(t)}p_ug_1^2\mathbb{E}[\mathrm{tr}\{(\mathbf{H}^T\mathbf{H}^*)^{-1}\}]\notag.$
Since $\mathbf{H}$ has i.i.d.~rows following $\mathcal{CN}(0,\mathbf{D})$, where $\mathbf{D}=\mathrm{diag}\{\beta_1, \beta_2, \cdots, \beta_K\}$, $\mathbf{H}^T\mathbf{H}^*$ is a $K\times K$ central Wishart matrix of $N$ degrees of freedom. Therefore, according to the properties of inverse Wishart matrices \cite{maiwald1997moments},  we have $\mathbb{E}\left[\left((\mathbf{H}^T\mathbf{H}^*)^{-1}\right)_{ii}\right]=\frac{1}{(N-K)\beta_i}$. Thus, 
\begin{align}
c_1&\approx \frac{p_u\alpha^{(t)}
}{N-K}	g_1^2 \sum_{i=1}^{K}\frac{1}{\beta_i}.\label{eq:c_1_2}
\end{align}  
The next is to calculate $c_2$. After using the $\mathbf{G}^{(t)}$ expression in (\ref{equation:3.16}),
\begin{align}
\begin{split}
c_2&=\alpha^{(t)}\mathrm{tr}\{\mathbf{G}_b^2\mathbb{E}[\mathbf{B}]\}\label{Eq:c_2_2}, 
\end{split}
\end{align}
where 
\begin{align}
\mathbf{B}\triangleq \mathbf{\tilde{H}}(\mathbf{\tilde{H}}^H\mathbf{\tilde{H}})^{-1}  \bD^{-\frac{1}{2}}
(\mathbf{P}^t)^T\bD^{-\frac{1}{2}}(\mathbf{\tilde{H}}^T\mathbf{\tilde{H}}^{*})^{-1}\bD^{-\frac{1}{2}}\mathbf{P}^t\bD^{-\frac{1}{2}}(\mathbf{\tilde{H}}^H\mathbf{\tilde{H}})^{-1}\mathbf{\tilde{H}}^H.\notag
\end{align}
Consider the singular-value decomposition (SVD)
\begin{align}
\mathbf{\tilde{H}}=\mathbf{U}\mathbf{\Sigma}\mathbf{V}^H,\label{eq:sing_val_dec} 
\end{align}
where $\mathbf{U}$, $\mathbf{V}$, and $\mathbf{\Sigma}$  are $N\times K$, $K \times K$, and $K\times K$ matrices. $\mathbf{U}$ and $\mathbf{V}$ contain singular vectors of $\mathbf{\tilde{H}}$ and $\mathbf{\Sigma}=\mathrm{diag}\{\sigma_1, \sigma_2,\cdots, \sigma_K\}$ contains the singular-values of $\mathbf{\tilde{H}}$. Further, according to  Definition 2.5 in \cite{tulino2004random}, $\mathbf{U}$, and $\mathbf{V}$ are Haar (isotropically) distributed matrices. We have
\begin{align*}
\mathbf{B}=\mathbf{U}\mathbf{\Sigma}^{-1}\mathbf{V}^H\mathbf{D}^{-\frac{1}{2}}(\mathbf{P}^t)^T\mathbf{D}^{-\frac{1}{2}}\mathbf{V}^*\mathbf{\Sigma}^{-2}\mathbf{V}^T\mathbf{D}^{-\frac{1}{2}}\mathbf{P}^t\mathbf{D}^{-\frac{1}{2}}
\mathbf{V}\mathbf{\Sigma}^{-1}\mathbf{U}^H.
\end{align*}
Let $\mathbf{P}_{ij}$ be the unitary permutation matrix that changes the positions of the $i$-th and the $j$-th rows of a matrix if it is multiplied from the left side. Then, using the fact that $\mathbf{U}$ and $\mathbf{P}_{ij}\mathbf{U}$ have the same distribution, we conclude that  
\begin{align*}
\mathbf{B}^\prime\triangleq \mathbf{P}_{ij} \mathbf{B} \mathbf{P}_{ij}^H = \mathbf{P}_{ij}\mathbf{U}\mathbf{\Sigma}^{-1}\mathbf{V}^H\mathbf{D}^{-\frac{1}{2}}
(\mathbf{P}^t)^T\mathbf{D}^{-\frac{1}{2}}\mathbf{V}^*\mathbf{\Sigma}^{-2}\mathbf{V}^T\mathbf{D}^{-\frac{1}{2}}\mathbf{P}^t\mathbf{D}^{-\frac{1}{2}}\mathbf{V}\mathbf{\Sigma}^{-1}
        (\mathbf{P}_{ij}\mathbf{U})^H,
        \end{align*}
has the same distribution as $\mathbf{B}$. From the construction of ${\bf B'}$, we have $b_{ii}=b'_{jj}$. It can thus be concluded that $\mathbb{E}\{b_{ii}\}$ is the same for all $i\in \{1, 2, \cdots, N\}$ and (\ref{Eq:c_2_2}) can be written as:
\begin{align}
c_2=&\alpha^{(t)}\left(\frac{1}{N}\sum_{n=1}^{N}G_{b_n}^2\right)
\mathbb{E}[\mathrm{tr}\{\mathbf{B}\}].\label{eq:c_2}
\end{align}
Next, we calculate $\mathrm{tr}\{\mathbf{B}\}$.
\begin{align}
\mathrm{tr}\{\mathbb{E}[\mathbf{B}]\}=&\sum_{m=1}^{K}\frac{1}{{\beta_m\beta_{i(m)}}}{\sum_{k_1=1}^{K}\sum_{k_2=1}^{K}\mathbb{E}\left[\frac{\lvert v_{i(m)k_1}\rvert^2\lvert v_{mk_2}\rvert^2}{\sigma_{k_1}^2\sigma_{k_2}^2}\right]}.
\label{eq:c_1_B}
\end{align} 
According to Lemma \ref{lem:moments}, for $k_1 \ne k_2$ and any $m$ we have:
\begin{align}
\mathbb{E}[\lvert v_{i(m)k_1}\rvert^2\lvert v_{mk_2}\rvert^2]=\frac{1}{K^2-1}.\label{eq:c_2_1}
\end{align}
Moreover, according to Theorem 1 in \cite{martin2004asymptotic}, the eigenvalues of the Wishart matrix $\mathbf{\tilde{H}}^H\mathbf{\tilde{H}}$, which are $\{\sigma_1^2, \sigma_2^2, \cdots , \sigma_K^2\}$, become independent as $N\rightarrow \infty$. Thus, when $k_1 \ne k_2$ and $N \gg 1$,
\begin{align}
\mathbb{E}\left[\frac{1}{\sigma_{k_1}^2\sigma_{k_2}^2}\right]\approx\left(\mathbb{E}\left[\frac{1}{\sigma^2_{k_1}}\right]\right)^2=\frac{1}{(N-K)^2},\label{eq:c_2_2}
\end{align}
where for the last step we have used the equality $\mathbb{E}\left[\frac{1}{\sigma_k^2}\right]=\frac{1}{K}\mathbb{E}[{\mathrm{tr}\{(\mathbf{\tilde{H}}^H\mathbf{\tilde{H}})^{-1}\}}]=\frac{1}{N-K}$, for any $k$. Also, since entries of $\mathbf{\tilde{H}}$ follow i.i.d. $\mathcal{CN}(0,1)$,  $\mathbf{U}$, $\mathbf{V}$, and $\mathbf{\Sigma}$ are independent. By using (\ref{eq:c_2_1}) and (\ref{eq:c_2_2}) in (\ref{eq:c_1_B}), for $k_1\ne k_2$,
\begin{align}
\mathbb{E}\left[\frac{\lvert v_{i(m)k_1}\rvert^2\lvert v_{mk_2}\rvert^2}{\sigma_{k_1}^2\sigma_{k_2}^2}\right]=\mathbb{E}[\lvert v_{i(m)k_1}\rvert^2\lvert v_{mk_2}\rvert^2]\mathbb{E}\left[\frac{1}{\sigma_{k_2}^2\sigma_{k_1}^2}\right]\approx\frac{1}{(K^2-1)(N-K)^2}.\label{eq:c_2_part1}
\end{align}
Also, according to Lemma \ref{lem:moments}, for $k_1=k_2=k$, and any $m$ and $k$ we have:
\begin{align}
\mathbb{E}[\lvert v_{i(m)k}\rvert^2\lvert v_{mk}\rvert^2]=\frac{1}{K(K+1)}.\label{eq:c_2_3}
\end{align}
Moreover, for any $k$,
\begin{align}
\mathbb{E}\left[\frac{1}{\sigma_{k}^4}\right]&=\frac{1}{K}\mathbb{E}\left[\mathrm{tr}\left\{(\mathbf{\tilde{H}}^H\mathbf{\tilde{H}})^{-1}\left((\mathbf{\tilde{H}}^H\mathbf{\tilde{H}})^{-1}\right)^H\right\}\right]\notag=\frac{1}{K}\mathbb{E}[{\mathrm{tr}\{(\mathbf{\tilde{H}}^H\mathbf{\tilde{H}})^{-2}\}}]\notag\\
&=\frac{1}{(N-K)(N-K-1)},\label{eq:c_2_4}
\end{align}  
where the last step is from results of the second order statistics of the inverse Wishart matrix in \cite{maiwald1997moments}. By combining (\ref{eq:c_2_3}) and (\ref{eq:c_2_4}), for $k_1=k_2=k$,
\begin{align}
\mathbb{E}\left[\frac{\lvert v_{i(m)k}\rvert^2\lvert v_{mk}\rvert^2}{\sigma_{k}^4}\right]=\frac{1}{K(K+1)(N-K)(N-K-1)}.\label{eq:c_2_part2}
\end{align}
By using (\ref{eq:c_2_part1}) and (\ref{eq:c_2_part2}) in  (\ref{eq:c_1_B}) and then (\ref{eq:c_2}), we have
\begin{align}
\begin{split}
c_2&\approx\frac{\alpha^{(t)}g_2(NK+N-2K-K^2)}{(N-K)^2(N-K-1)(K+1)}\sum_{m=1}^{K}\frac{1}{\beta_m\beta_{i(m)}}\end{split}.\label{Eq:c_2_done}
\end{align}
For $c_3$, with similar arguments as $c_2$, we can show that
\begin{align}
\begin{split}
c_3=&\mathrm{tr}\{\mathbb{E}[\mathbf{d}\mathbf{d}^H]\mathbb{E}[(\mathbf{G}^{(t)})^H\mathbf{G}^{(t)}]\}= \mathrm{tr}\{\mathbf{C}_\mathbf{d}\mathbb{E}[(\mathbf{G}^{(t)})^H\mathbf{G}^{(t)}]\}
\\
\approx&\frac{\alpha^{(t)}(\hat{c}-v g_2)(NK+N-2K-K^2)}{(N-K)^2(N-K-1)(K+1)}\sum_{m=1}^{K}\frac{1}{\beta_m\beta_{i(m)}}\label{Eq:c_3_2}.
\end{split}
\end{align}
The details are omitted due to the page limit. By combining (\ref{eq:c_1_2}), (\ref{Eq:c_2_done}), and (\ref{Eq:c_3_2}) and also ignoring lower order terms of $N$ for large $N$, i.e., ($NK\gg K$ when $N\gg1$), (\ref{eq:p_zf_coeff}) is obtained.
\section{Proof of {\color{black}Theorem} \ref{theo:Theorem_ach}} \label{App:proof_rate}
We use the common approximation $\mathbb{E}\{\log_2(1+\frac{X}{Y})\}\approx\log_2(1+\frac{\mathbb{E}\{X\}}{\mathbb{E}\{Y\}})$ for massive MIMO systems. It is tight when $N\rightarrow\infty$ and $X$ and $Y$ are both sums of nonnegative random variables which converge to their means due to the law of large numbers \cite{zhang2014power}. Therefore,
\begin{align}
R_{k,i(k)}&\approx \log_2\left(1+\frac{A_4}{A_1+A_2+A_3+1}\right),\notag
\end{align}
where
\begin{align}
A_1\triangleq&p_u\mathbb{E}\left[\sum_{j=1, j\ne i(k) }^{K}{\abs{\mathbf{h}_{k}^{T}\mathbf{G}^{(t)}\mathbf{G}_b\mathbf{h}_j}^2}\right]\notag, \hspace{0.3 cm}
A_2\triangleq\mathbb{E}[\norm{{\mathbf{h}_{k}}^{T}\mathbf{G}^{(t)}\mathbf{G}_b}^2]\notag,\\
A_3\triangleq&\mathbb{E}[\norm{{\mathbf{h}_{k}}^{T}\mathbf{G}^{(t)}\mathbf{d}}^2]\notag,\hspace{0.3 cm} 
A_4\triangleq p_u\mathbb{E}[ \abs{{\mathbf{h}_{k}}^{T}\mathbf{G}^{(t)}\mathbf{G}_b\mathbf{h}_{i(k)}}^2]\notag.
\end{align}
 {\color{black}Similar to the proof in Appendix \ref{App:proof_p_zf}, in calculating $A_1, A_2, A_3$, and $A_4$, we first simplify the expressions by the SVD of the channel matrix. Then, we use properties of Wishart distributed matrices, Haar distributed matrices, and the singular values for Gaussian matrix to calculate the values.}

Let $\mathbf{e}_k$ {\color{black}be} the $k$th canonical basis. Substituting $\mathbf{G}^{(t)}$ from (\ref{equation:3.16}) in $A_1$, we have
\begin{align}
A_1 &= p_u\alpha^{(\hspace{-0.5mm}t\hspace{-0.5mm})} \hspace{-4mm}
\sum_{j=1, j\ne i(k)}^{K} \hspace{-4mm}
\mathbb{E}\hspace{-1mm}\left[\hspace{-0.5mm}\mathbf{e}_{i(k)}^T \hspace{-0.5mm}
(\hspace{-0.5mm}\mathbf{H}^{\hspace{-0.5mm}H}\mathbf{H})^{\hspace{-.5mm}-\hspace{-.5mm}1}\mathbf{H}^H\mathbf{G}_b\mathbf{h}_j\mathbf{h}_j^H\mathbf{G}_b^H
\mathbf{H}(\hspace{-0.5mm}\mathbf{H}^{\hspace{-0.5mm}H}\mathbf{H}\hspace{-0.5mm})^{\hspace{-0.5mm}-\hspace{-0.5mm}1}
\mathbf{e}_{i(\hspace{-0.5mm}k\hspace{-0.5mm})}\hspace{-0.5mm}\right] \nonumber\\ &=\frac{p_u\alpha^{(t)}}{\beta_{i(k)}} \hspace{-2mm}
\sum_{j=1, j\ne i(k)}^{K} \hspace{-4mm}
{\beta_j}\mathbb{E}\left[\left|
((\mathbf{\tilde{H}}^H\mathbf{\tilde{H}})^{-1}\mathbf{\tilde{H}}^H
\mathbf{G}_b\mathbf{\tilde{H}})_{i(k)j}\right|^2\right].\notag
\end{align}
By using the SVD in (\ref{eq:sing_val_dec}),
\begin{align}
A_1&=\frac{p_u\alpha^{(t)}}{\beta_{i(k)}} \hspace{-2mm}
\sum_{j=1, j\ne i(k)}^{K} \hspace{-4mm}
{\beta_j}\mathbb{E}\left[
\left|(\mathbf{V}\mathbf{\Sigma}^{-1}\mathbf{U}^H\mathbf{G}_b\mathbf{U}\mathbf{\Sigma}\mathbf{V}^H)_{i(k)j}\right|^2\right] \notag  \\
&=\frac{p_u\alpha^{(t)}}{\beta_{i(k)}} \hspace{-2mm}
\sum_{j=1, j\ne i(k)}^{K} \hspace{-4mm} {\beta_j}
\mathbb{E}\hspace{-1.5mm}\left[\hspace{-0.5mm}\left|\sum_{k_1=1}^{K}\sum_{k_2=1}^{K}
\hspace{-1mm}\left(\hspace{-1mm}\frac{\sigma_{k_2}}{\sigma_{k_1}}\mathbf{v}_{k_1}
\mathbf{u}_{k_1}^H\mathbf{G}_b\mathbf{u}_{k_2}\mathbf{v}_{k_2}^H\hspace{-1mm}
\right)_{\hspace{-1mm}i(k)j}\right|^2\hspace{-0.5mm}\right]\notag\\
&=\frac{p_u\alpha^{(t)}}{\beta_{i(k)}} \hspace{-0.3cm}
\sum_{j=1, j\ne i(k)}^{K} \hspace{-4mm} {\beta_j}
\sum_{k_1,k_2}\sum_{k'_1,k'_2}\mathbb{E}\bigg[\frac{\sigma_{k_2}\sigma_{k'_2}}{\sigma_{k_1}\sigma_{k'_1}}
v_{i(k),k_1}v_{j,k_2}^*v_{i(k),k'_1}^*v_{j,k'_2}
 \mathbf{u}_{k_1}^H\mathbf{G}_b\mathbf{u}_{k_2}\mathbf{u}_{k'_2}^H\mathbf{G}_b\mathbf{u}_{k'_1}\bigg].\label{Eq:complex_A_2}
\end{align}
As mentioned before, $\mathbf{U}$, $\mathbf{V}$, and $\mathbf{\Sigma}$ are independent. \textcolor{black}{Thus}, from Lemma \ref{lem:moments}, if at least one of $k_1, k_2, k'_1, k'_2 $ is different from the others, the corresponding expectation term in (\ref{Eq:complex_A_2}) is $0$. The remaining terms in the summation in (\ref{Eq:complex_A_2}), are considered in the following four cases.
{\color{black}\begin{enumerate}
\item If $k_1=k_2$ , $k'_1=k'_2$, and $k_1\ne k'_1$,
        \begin{align}
        \begin{split}
        b_1\triangleq&\sum_{k_1=1}^{K}\sum_{k'_1=1, \ne k_1 }^{K}\mathbb{E}\bigg[v_{i(k),k_1}v_{j,k_1}^*v_{i(k),k'_1}^*v_{j,k'_1}\mathbf{u}_{k_1}^H\mathbf{G}_b\mathbf{u}_{k_1}\mathbf{u}_{k'_1}^H\mathbf{G}_b\mathbf{u}_{k'_1}\bigg]\notag\\
=&-\frac{1}{(K+1)} \left[\frac{\sum_{n=1}^{N}G_{b_n}^2}{N(N+1)}+\frac{\sum_{n_1=1}^{N}\sum_{n_2=1, \ne n_1}^{N}G_{b_{n_1}}G_{b_{n_2}}}{N^2-1}\right].
        \end{split}
        \end{align}
\item If $k_1=k'_1$, $k_2=k'_2$, and $k_1\ne k_2$,
        \begin{align}
        \begin{split}
        b_2\triangleq&\sum_{k_1=1}^{K}\sum_{k_2=1,\ne k_1 }^K\mathbb{E}\left[
\frac{\sigma_{k_2}^{2}}{\sigma_{k_1}^2}
|v_{i(k),k_1}|^2|v_{j,k_2}|^2|\mathbf{u}_{k_1}^H\mathbf{G}_b\mathbf{u}_{k_2}|^2\right]\\
        =&\sum_{k_1=1}^{K}\sum_{k_2=1, \ne k_1 }^{K}\mathbb{E}\left[\frac{\sigma_{k_2}^{2}}{\sigma_{k_1}^{2}}\right]\frac{1}{(K^2-1)}\left[\frac{\sum_{n=1}^{N}G_{b_n}^2}{N(N+1)}-\frac{\sum_{n_1=1}^{N}\sum_{n_2=1,\ne n_1}^{N}G_{b_{n_1}}G_{b_{n_2}}}{N(N^2-1)}\right].\notag
        \end{split}
        \end{align}
        As mentioned earlier in the proof of Theorem \ref{lem:alpha}, for $k_1\ne k_2$, $\sigma_{k_1}^2$ and $\sigma_{k_2}^2$ are unordered eigenvalues of Wishart matrix which become independent as $N\rightarrow \infty$. Thus, for $N\gg 1,$
        \begin{align*}
        \sum_{k_1=1}^{K}\sum_{k_2=1,\ne k_1}^{K}\mathbb{E}\left[\frac{\sigma_{k_2}^2}{\sigma_{k_1}^2}\right]\approx\sum_{k_1=1}^{K}\sum_{k_2=1,\ne k_1}^{K}{\mathbb{E}[\sigma_{k_2}^2]}\mathbb{E}\left[\frac{1}{\sigma_{k_1}^2}\right].
        \end{align*}
  For any $k_1\ne k_2$, using properties of Wishart matrix, we have $\mathbb{E}\left[\frac{1}{\sigma_{k_1}^2}\right]=\frac{1}{N-K}$ and $   \mathbb{E}[\sigma_{k_2}^2]=\frac{1}{K}\mathbb{E}[
  \mathrm{tr}\{\mathbf{\tilde{H}}^H\mathbf{\tilde{H}}\}]=N$. Therefore,
    \begin{align*}
  b_2\approx\frac{NK}{(N-K)(K+1)}\left[\frac{\sum_{n=1}^{N}G_{b_n}^2}{N(N+1)}-\frac{\sum_{n_1=1}^{N}\sum_{n_2=1,\ne n_1}^{N}G_{b_{n_1}}G_{b_{n_2}}}{N(N^2-1)}\right].
  \end{align*}
\item If $k_1=k'_2$, $k_2=k'_1$, and $k_1\ne k_2$,
\begin{align*}
\begin{split}
b_3\triangleq\sum_{k_1=1}^{K}\sum_{k'_1=1, \ne k_1}^{K}\mathbb{E}\bigg[v_{i(k),k_1}v_{j,k'_1}^*v_{i(k),k'_1}^*v_{j,k_1}\mathbf{u}_{k_1}^H\mathbf{G}_b\mathbf{u}_{k'_1}\mathbf{u}_{k_1}^H\mathbf{G}_b\mathbf{u}_{k'_1}\bigg]=0.
\end{split}
\end{align*}   
 \item If $k_1=k_2=k'_1=k'_2$,
   \begin{align*}
        b_4\triangleq\sum_{k=1}^{K}\mathbb{E}\left[|v_{i(k),k}|^2|v_{j,k}|^2|\mathbf{u}_k^H\mathbf{G}_b\mathbf{u}_k|^2\right]=\left[\frac{\sum_{n=1}^{N}G_{b_n}^2+\sum_{n_1=1}^{N}\sum_{n_2=1}^{N}G_{b_{n_1}}G_{b_{n_2}}}{(K+1)N(N+1)}\right].
\end{align*}
By using the above results on $b_1,b_2,b_3$, and $b_4$ in (\ref{Eq:complex_A_2}),  for $N\gg1$, 
\begin{align}
A_1&\approx\frac{p_u\alpha^{(t)}(N-K+NK)}{\beta_{i(k)}(K+1)(N-K)N(N^2-1)}\bigg[{N\sum_{n=1}^{N}\hspace{-0.5mm}
G_{b_n}^2}\hspace{-0.5mm}-\hspace{-1.5mm}\sum_{n_1=1}^{N}\hspace{-0.5mm}
\sum_{n_2=1}^{N}\hspace{-0.5mm}G_{b_{n_1}}\hspace{-0.5mm}G_{b_{n_2}}\bigg]
\hspace{-1mm}\sum_{j=1, j\ne i(k) }^{K}\hspace{-2.5mm}\beta_j \nonumber\\
&\approx \frac{p_u}{\beta_{i(k)}}\frac{\alpha^{(t)}}{(N-K)}(g_2-g_1^2)
\beta_{\backslash i(k)}.
\label{Eq:A_1}
\end{align}
\end{enumerate}}
Next, we calculate $A_2$. After substituting $\mathbf{G}^{(t)}$ from (\ref{equation:3.16}), 
\begin{align}
A_2=&\alpha^{(t)}\mathbb{E}\left[\mathbf{e}_{i(k)}^T\left(\mathbf{H}^H\mathbf{H}\right)^{-1}\mathbf{H}^H\mathbf{G}_b^2\mathbf{H}\left(\mathbf{H}^H\mathbf{H}\right)^{-1}\mathbf{e}_{i(k)}\right]\notag\\
=&\frac{\alpha^{(t)}}{\beta_{i(k)}}\mathbb{E}\left[\left(\left(\mathbf{\tilde{H}}^H\mathbf{\tilde{H}}\right)^{-1}\mathbf{\tilde{H}}^H\mathbf{G}_b^2\mathbf{\tilde{H}}\left(\mathbf{\tilde{H}}^H\mathbf{\tilde{H}}\right)^{-1}\right)_{i(k)i(k)}\right]\notag\\
=&\frac{\alpha^{(t)}}{\beta_{i(k)}}\mathbb{E}\left[\left(\mathbf{V}\mathbf{\Sigma}^{-1}\mathbf{U}^H\mathbf{G}_b^2\mathbf{U}\mathbf{\Sigma}^{-1}\mathbf{V}^H
\right)_{i(k)i(k)}\right].\notag
\end{align}
Following similar reasoning as the one which led to (\ref{eq:c_2}), we have
\begin{align}
A_2&=\frac{\alpha^{(t)}}{\beta_{i(k)}K}\mathbb{E}\left[\mathrm{tr}(\mathbf{G}_b^2\mathbf{U}\mathbf{\Sigma}^{-2}\mathbf{U}^H)\right]=\frac{\alpha^{(t)}}{\beta_{i(k)}K}\mathbb{E}\left[\mathrm{tr}(\mathbf{G}_b^2\sum_{k=1}^{K}\sigma_k^{-2}\mathbf{u}_k\mathbf{u}_k^H)\right].\label{Eq:assym_simp_A3}
\end{align} 
From Lemma \ref{lem:moments}, $\mathbb{E}[u_{ik}u_{jk}^*]=0$ for $i\ne j$ and $\mathbb{E}[|u_{nk}|^2]=\frac{1}{N}$ for all $n,k$. So, from (\ref{Eq:assym_simp_A3}),
\begin{align}
A_2&=\frac{\alpha^{(t)}}{\beta_{i(k)}K}\mathbb{E}\left[\sum_{k=1}^{K}\sigma_k^{-2}\sum_{n=1}^{N}G_{b_n}^2|u_{nk}|^2\right]=\frac{\alpha^{(t)}}{\beta_{i(k)}K}\mathbb{E}\left[\sum_{k=1}^{K}\sigma_k^{-2}\right]\left(\frac{1}{N}\sum_{n=1}^{N}G_{b_n}^2\right).\notag
\end{align}
As mentioned earlier, for the inverse Wishart matrix $(\mathbf{\tilde{H}}^H\mathbf{\tilde{H}})^{-1}$, we have  $\mathbb{E}\left[\sum_{k=1}^{K}\frac{1}{\sigma_k^2}\right]=\\\mathbb{E}\left[{\mathrm{tr}\{(\mathbf{\tilde{H}}^H\mathbf{\tilde{H}})^{-1}\}}\right]=\frac{K}{N-K}$. Thus,
\begin{align}
A_2=\frac{1}{\beta_{i(k)}}
\frac{\alpha^{(t)}}{N-K}g_2. \label{Eq:A_2}
\end{align}
Similarly, $A_3$ can be found as 
\begin{align}
A_3&=\frac{1}{\beta_{i(k)}}\frac{\alpha^{(t)}}{N-K}
\left(\hat{c}-v g_2\right).
\label{Eq:A_3}
\end{align}
Finally, after substituting $\mathbf{G}^{(t)}$ from (\ref{equation:3.16}) in $A_4$, we have 
\begin{align}
A_4=& {p_u\alpha^{(t)}}\mathbb{E}\left[\left|\left(\left(\mathbf{\tilde{H}}^H\mathbf{\tilde{H}}\right)^{-1}\mathbf{\tilde{H}}^H\mathbf{G}_b\mathbf{\tilde{H}}\right)_{i(k)i(k)}\right|^2\right]\notag\\
=& {p_u\alpha^{(t)}}\sum_{k_1,k_2}\sum_{k'_1,k'_2}
\mathbb{E}\bigg[\frac{\sigma_{k_2}\sigma_{k'_2}}{\sigma_{k_1}\sigma_{k'_1}}v_{i(k),k_1}v_{i(k),k_2}^*v_{i(k),k'_1}^*
\label{Eq:complex_A_1}v_{i(k),k'_2} \mathbf{u}_{k_1}^H\mathbf{G}_b\mathbf{u}_{k_2}\mathbf{u}_{k'_2}^H\mathbf{G}_b\mathbf{u}_{k'_1}\bigg].
\end{align}
Similar to the derivations for $A_1$ and according to Lemma \ref{lem:moments}, if at least one of $k_1, k_2, k'_1, k'_2 $ is different from the others, the corresponding expectation term in (\ref{Eq:complex_A_1}) is $0$. The remaining terms are considered in the following four cases. 
{\color{black}\begin{enumerate}
	\item If $k_1=k_2$, $k'_1=k'_2$, and $k_1\ne k'_1$, the sum of the corresponding terms in (\ref{Eq:complex_A_1}) can be calculated as follows.
	\begin{align}
	\begin{split}
	d_1&=\frac{p_u\alpha^{(t)}(K-1)}{(K+1)}\bigg[\frac{\sum_{n=1}^{N}G_{b_n}^2}{N(N+1)}+\frac{\sum_{n_1=1}^{N}\sum_{n_2=1, \ne n_1}^{N}G_{b_{n_1}}G_{b_{n_2}}}{N^2-1}\bigg].\notag
	\end{split}
	\end{align}
	\item If $k_1=k'_1$, $k_2=k'_2$, and $k_1\ne k_2$, the sum of the corresponding terms in (\ref{Eq:complex_A_1}) can be calculated as follows.
	\begin{align}
	\begin{split}
	d_2&=\frac{p_u\alpha^{(t)}N(K-1)}{(K+1)(N-K)}\left[\frac{\sum_{n=1}^{N}G_{b_n}^2}{N(N+1)}-\frac{\sum_{n_1=1}^{N}\sum_{n_2=1,\ne n_1}^{N}G_{b_{n_1}}G_{b_{n_2}}}{N(N^2-1)}\right].\notag
	\end{split}
	\end{align}
	\item If $k_1=k'_2$, $k_2=k'_1$, and $k_1\ne k_2$,  each of the corresponding term in (\ref{Eq:complex_A_1}) is $0$. 
 	\item If $k_1=k_2=k'_1=k'_2$, the sum of the corresponding terms in (\ref{Eq:complex_A_1}) can be calculated as follows.
	\begin{align}
	\begin{split}
	& d_3=\frac{2p_u\alpha^{(t)}}{(K+1)}\left[\frac{\sum_{n=1}^{N}G_{b_n}^2+\sum_{n_1=1}^{N}\sum_{n_2=1}^{N}G_{b_{n_1}}G_{b_{n_2}}}{N(N+1)}\right].\notag
	\end{split}
	\end{align}
\end{enumerate}}
\textcolor{black}{Hence}, using the above results of $d_1, d_2, d_3$ in (\ref{Eq:complex_A_1}),  for $N\gg1$,
\begin{align}
A_4=&\frac{p_u\alpha^{(t)}}{(K+1)(N-K)(N^2-1)}\times \bigg[{(N^2K+N^2-N-3NK+K+K^2)g_2}\notag\\
&\hspace{5cm}+N(N^2K+N^2-NK^2-N-2NK+2K)g_1^2\bigg] \notag \\
\approx&  p_u\frac{\alpha^{(t)}}{N-K}
\bigg[\frac{Kg_2+(NK+N-K^2-2K)g_1^2}{K+1}\bigg]. \label{Eq:A_4}
\end{align}
By using (\ref{Eq:A_1}), (\ref{Eq:A_2}), (\ref{Eq:A_3}), (\ref{Eq:A_4}), {\color{black}and ignoring lower order terms of $N$ for large $N$, ( e.g., $NK\gg K$ when $N\gg1$)} the average achievable rate result in {\color{black}Theorem} \ref{theo:Theorem_ach} is obtained.

        \bibliographystyle{IEEEtran}
        \bibliography{IEEEexample}
        %
        %
        %

\end{document}